\documentclass{amsart}
\usepackage{epsfig}
\usepackage{hyperref}
\usepackage{verbatim,amssymb,amsfonts,amsthm,amsmath,amscd,ifthen}
\def\sgn{\qopname \relax o{sgn}}
{     \theoremstyle{plain}
        \newtheorem{theorem}{Theorem}[section]
          \newtheorem{corollary}[theorem]{Corollary} 
          \newtheorem{proposition}[theorem]{Proposition} 
}
{       \theoremstyle{remark}
        \newtheorem{remark}{Remark}[section]
}
\makeatletter
\@addtoreset{equation}{section}
\makeatother
\begin{document}
\def\spIon{Q_{1,1}} 
\def\spIIon{Q_{1,2}} 
\def\spIIIon{Q_{1,3}} 
\def\spItt{Q_{2,1}} 
\def\spIItt{Q_{2,2}} 
\def\spIIItt{Q_{2,3}} 
\def\spJtop{Q_{3,1}} 
\def\spJtopp{Q_{3,2}} 
\def\spItf{Q_{4,1}} 
\def\spIItf{Q_{4,2}} 
\def\spIIItf{Q_{4,3}} 
\def\spIte{Q_{5,1}} 
\def\spIIte{Q_{5,2}} 
\def\spIfz{Q_{6,1}} 
\def\spIIfz{Q_{6,2}} 
\def\spIIIfz{Q_{6,3}} 
\def\spIst{Q_{7,1}} 
\def\spIIst{Q_{7,2}} 
\def\specIthree{Q_{8,1}} 
\def\specIIthree{Q_{8,2}} 
\def\specIIIthree{Q_{8,3}} 
\def\specIsix{Q_{9,1}} 
\def\specIIsix{Q_{9,2}} 
\def\specIIIsix{Q_{9,3}} 
\def\specIseven{Q_{10,1}} 
\def\specIIseven{Q_{10,2}} 
\def\specIeight{Q_{11,1}} 
\def\specIIeight{Q_{11,2}} 
\def\specIIIeight{Q_{11,3}} 
\def\too{{\dot{\to}}}
\def\trace{\mathop{\rm Trace}}
\def\N{\mathbb{N}}
\def\R{\mathbb{R}}
\def\E{\mathbb{E}}
\def\ls{\left[}
\def\ZP{{\mathbb{Z}}_{\geq0}}
\def\Z{{\mathbb{Z}}}
\def\C{{\mathbb{C}}}
\def\rs{\right]}\def\lp{\left(}\def\rp{\right)}
\def\diag{\qopname \relax o{diag}}
\def\vol{\qopname \relax o{vol}}
\def\Tr{\qopname \relax o{Tr}}
\def\tr{\qopname \relax o{tr}}
\def\dist{\qopname \relax o{dist}}
\def\dom{\qopname \relax o{Dom}}
\def\Prob{\qopname \relax o{Prob}}
\def\spec{\qopname \relax o{spec}}
\def\Ai{\qopname \relax o{Ai}}
\def\Aip{\Ai^\prime}
\def\xin{\xi^{(N)}}
\def\etan{\eta^{(N)}}
\def\ra{\rightarrow}
\def\eps{\epsilon}
\def\sgn{\qopname \relax o{sgn}}
\def\KA{K_{{\textrm{Airy}}}}
\def\const{{\rm const}\relax }
\def\sno{S_N^{(1)}}
\def\snop{S_N^{(1)\prime}}
\def\snf{S_N^{(4)}}
\def\snt{S_N^{(2)}}
\def\ph{\phi}
\def\corrf{\qopname \relax o{Corr}_N^{(4)}}
\def\corro{\qopname \relax o{Corr}_N^{(1)}}
\newcommand{\nc}{\newcommand}
\nc{\onetwo}[2]{\left(\begin{array}{cc}#1&#2\end{array}\right)}
\nc{\twotwo}[4]{\left(\begin{array}{cc}#1&#2\\&\\#3&#4\end{array}\right)}
\nc{\twoone}[2]{\left(\begin{array}{c}#1\\\\#2\end{array}\right)}
\nc{\mtt}[4]{\begin{pmatrix}#1&#2\\#3&#4\end{pmatrix}}
\newcommand{\bp}{\begin{proof}}
\newcommand{\ep}{\end{proof}}
\newcommand{\bt}{\begin{theorem}}
\newcommand{\et}{\end{theorem}}
\newcommand{\be}{\begin{equation}}
\newcommand{\ee}{\end{equation}}
\newcommand{\bq}{\begin{equation}}
\newcommand{\eq}{\end{equation}}
\newcommand{\ba}{\begin{aligned}}
\newcommand{\ea}{\end{aligned}}
\newcommand{\la}[1]{\label{#1}}
\nc{\fh}{\hat{f}}
\nc{\Fh}{\hat{F}}
\nc{\ft}{\tilde{f}}
\nc{\fth}{\hat{\tilde{f}}}
\nc{\pht}{\tilde{\phi}}
\nc{\pth}{\hat{\tilde{\phi}}}
\nc{\cd}{\cdots}
\nc{\ov}{\over}
\newcommand{\er}{\eqref}

\title[Universality at the edge for invariant
ensembles]
{Universality at the edge of the spectrum
 for unitary, orthogonal and symplectic ensembles of random matrices}

\author[Deift and Gioev]
{Percy Deift and Dimitri Gioev}
\address{Deift: Department of Mathematics, Courant Institute of Mathematical
Sciences, New York University, New York, NY 10012}
\email{deift@cims.nyu.edu}
\address{Gioev:
Department of Mathematics, Courant Institute of Mathematical
Sciences, New York University, New York, NY 10012,
and Department of Mathematics, University of Rochester, Rochester, NY 14627}
\email{gioev@cims.nyu.edu}
\begin{abstract}
We prove universality at the edge of the spectrum
for unitary ($\beta=2$), orthogonal ($\beta=1$)
and symplectic ($\beta=4$) ensembles of random matrices
in the scaling limit
for a class of weights $w(x)=e^{-V(x)}$ where $V$ is a polynomial,
$V(x)=\kappa_{2m}x^{2m}+\cdots$, $\kappa_{2m}>0$.
The precise statement of our results is given in Theorem~\ref{thmuniv}
and Corollaries \ref{cor1.1}, \ref{cor1.2} below.
For a proof of universality in the bulk of the spectrum,
 for the same class of weights,
for unitary ensembles
see \cite{DKMVZ2}, and for orthogonal
and symplectic ensembles see \cite{DG}.

Our starting point in the unitary case is \cite{DKMVZ2},
and for the orthogonal and symplectic cases we rely on
our recent work \cite{DG}, 
which in turn depends on the earlier work of
Widom \cite{W} and  Tracy and Widom \cite{TW2}.
As in \cite{DG}, the uniform Plancherel--Rotach type asymptotics
for the orthogonal polynomials 
found in \cite{DKMVZ2} plays a central role.

The formulae in \cite{W} express the correlation kernels for $\beta=1$ and $4$
as a sum of a Christoffel--Darboux (CD) term, as in the case $\beta=2$,
together with a correction term. In the bulk scaling limit \cite{DG},
the correction term is of lower order and does not contribute
to the limiting form of the correlation kernel.
By contrast, in the edge scaling limit considered here,
the CD term and the correction term contribute to the same order:
this leads to additional technical difficulties over and above \cite{DG}.
\end{abstract}
\maketitle
\section{Introduction}
\la{sec1}
This paper is a continuation of \cite{DG}. In \cite{DG}, the authors proved
universality in the bulk for orthogonal and symplectic ensembles: here we prove
universality at the edge for orthogonal and symplectic ensembles,
and also for unitary ensembles. For the convenience of the reader,
and to fix notation, we now summarize some of the basic theory
of invariant ensembles ($\beta=1$, $2$ or $4$),
borrowing freely and extensively from the introduction in \cite{DG}.
We are concerned with ensembles
of  matrices $\{M\}$ with probability distributions
\be
\la{eq1p1}
     \mathcal{P}_{N,\beta}(M)\,dM
  = \frac1{\mathcal{Z}_{N,\beta}}\,e^{-\tr V_\beta(M)}\,dM,
\ee
for $\beta=1$, $2$ and $4$, the so-called Orthogonal, Unitary
and Symplectic ensembles, respectively (see \cite{M}).
For $\beta=1$, $2$, $4$, the ensemble consists of $N\times{}N$
real symmetric matrices, $N\times{}N$
Hermitian matrices, and $2N\times{}2N$
Hermitian self-dual matrices, respectively.
In general the potential $V_\beta(x)$
is a real-valued function growing sufficiently rapidly as $|x|\to\infty$,
but we will restrict our attention henceforth to $V_\beta$'s
which are polynomials,
\be
\la{eq1p2}
   V_\beta(x)=\kappa_{2m,\beta}x^{2m}+\cdots,\qquad\kappa_{2m,\beta}>0.
\ee
In \eqref{eq1p1}, $dM$ denotes Lebesgue measure on the
algebraically independent
entries of $M$, and $\mathcal{Z}_{N,\beta}$ is a normalization constant.
The above terminology for $\beta=1$, $2$ and $4$
reflects the fact that \eqref{eq1p1}
is invariant under conjugation of $M$, $M\mapsto{}UMU^{-1}$,
by orthogonal, unitary and unitary-symplectic
matrices $U$.
It follows from \eqref{eq1p1} that the distribution of the
eigenvalues $x_1,\cdots,x_N$
of $M$ is given (see \cite{M}) by
\begin{equation}
\label{eq_PN}
     P_{N,\beta}(x_1,\cdots,x_N) =  \frac1{{Z}_{N,\beta}}
         \prod_{1\leq{}j<k\leq{}N}
        |x_j-x_k|^\beta\prod_{j=1}^N w_\beta(x_j)
\end{equation}
where again $Z_{N,\beta}$ is a normalization constant (partition function).
Here
\be
\la{eq1p3}
   w_\beta(x)=\begin{cases}
                  e^{-V_\beta(x)},&\beta=1,2\cr
                  e^{-2V_\beta(x)},&\beta=4.
               \end{cases}
\ee
(The factor $2$ in $w_{\beta=4}$ reflects the fact that the
eigenvalues of self-dual Hermitian matrices come in pairs.)
Let $\{p_j\}_{j\geq0}$ be the normalized orthogonal polynomials (OP's)
on $\R$ with respect to the weight $w\equiv{}w_{\beta=2}$,
and define $\phi_j\equiv{}p_j{}w^{1/2}$.
Note that  $(\phi_j,\phi_k)=\delta_{jk}$
where $(\cdot,\cdot)$ denotes the standard inner product in $L^2(\R)$.

For the unitary matrix ensembles 
an important role is played by the Christoffel--Darboux (CD)
kernel
\bq\label{K}
      K_N(x,y)\equiv K_{N,2}(x,y)=\sum_{k=0}^{N-1}\ph_k(x)\,\ph_k(y).
\eq
In particular the probability density \eqref{eq_PN},
the $l$-point correlation function $R_{N,l,2}$
and also the gap
probability $E_2(0;J)$ that a set $J$ contains no eigenvalues,
can all be expressed in terms of $K_N$, see e.g.~\cite{M}.
For example
\be\la{eq1a}
   R_{N,l,2}(x_1,\cdots,x_l)=\det(K_N(x_j,x_k))_{1\leq{}j,k\leq{}l}.
\ee

The Universality Conjecture, in our situation,
states that the limiting statistical behavior of the eigenvalues
$x_1,\cdots,x_N$ distributed according to the law \eqref{eq_PN},
in the appropriate scale as $N\ra\infty$,
should be independent of the weight $w_\beta$,
and should depend only on the invariance properties of $\mathcal{P}_{N,\beta}$,
$\beta=1$, $2$ or $4$, mentioned above.
Universality has been considered extensively in the physics literature,
see e.g.~\cite{BZ,Be,HW,SV}.

The kernel $K_N(x,y)$ can also be expressed
via the Christoffel--Darboux formula
\begin{equation}
\label{eq_KN}
     K_N(x,y)= b_{N-1}\,\frac{\phi_{N}(x)\phi_{N-1}(y)
         -\phi_{N-1}(x)\phi_{N}(y)}{x-y},
\end{equation}
where $b_{N-1}$ is a coefficient
in the three-term recurrence relation for OP's, see \cite{Sz}.
In view of the preceding remarks it follows that
in the case $\beta=2$,
the study of the large $N$ behavior of
$P_{N,2}$, and in particular the proof of universality,
reduces to the
asymptotic analysis of $b_{N-1}$ and the OP's $p_{N+j}$ with $j=0$ or $-1$.
By a fundamental observation
of Fokas, Its and Kitaev [FoIKi]
 the OP's solve
a Riemann--Hilbert problem (RHP) of a type
that is amenable to the steepest descent method
introduced by Deift and Zhou in \cite{DZ} and further developed in \cite{DVZ}.
In \cite{DKMVZ,DKMVZ2} the authors
analyzed the asymptotics of OP's for very general classes of weights.
In particular they proved the Universality Conjecture in the bulk
in the case $\beta=2$ for weights
$w(x)=e^{-V(x)}$ where $V(x)$ is a polynomial as above,
and also for $w(x)=e^{-NV(x)}$ where $V(x)$ is real analytic and
$V(x)/\log|x|\to+\infty$, as $|x|\to\infty$.
The bulk scaling limit as $N\to\infty$
is described in terms of the so-called {\em sine kernel\ }
$K_\infty(x-y)$ where
\be\la{eqsinek}
            K_\infty(t)\equiv\frac{\sin\pi t}{\pi t}.
\ee
For example \cite[Theorem~1.4]{DKMVZ2},
for $w(x)=e^{-V(x)}$, $V(x)$ polynomial,
and for any $l=2,3,\cdots$ and $r,y_1,\cdots,y_l$ in a compact
set, one has as $N\ra\infty$
\be\la{eq1b}
   \frac{1}{(K_N(0,0))^l}\,R_{N,l,2}\Big(r+\frac{y_1}{K_N(0,0)},
   \cdots,\,
r+\frac{y_l}{K_N(0,0)}\Big)
\ra\det(K_\infty(y_j-y_k))_{1\leq j,k\leq l}.
\ee
The scale $x=y/K_N(0,0)$ is chosen so that the expected number of
eigenvalues per unit $y$-interval is one.
 This scaling in the bulk
is standard in Random Matrix Theory.
Indeed for any Borel set $B\subset\R$,
\be\la{eqs3p1}
  \int_{B} R_{N,l=1,2}(x)\,dx = \E\{\,\textrm{number of eigenvalues in $B$}\,\}.
\ee
Thus by \eqref{eq1a} $K_N(0,0)=R_{N,1,2}(0)$ gives the density
of the expected number of eigenvalues near zero.
From \eqref{eq1b}, we see that, in the appropriate scale,
the large $N$ behavior of the eigenvalues is {\em universal\ }(i.e.~independent of $V$).
Pioneering mathematical work on the Universality Conjecture
in the bulk was done in \cite{PS}
and for the case of quartic two-interval potential $V(x)=N(x^4-tx^2)$,
$t>0$ (sufficiently) large, in \cite{BI}.
We note again that
 all these results apply only in the case $\beta=2$.

In the case $\beta=1$ and $4$ the situation is more complicated.
In place of \eqref{K} one must use
$2\times2$ matrix kernels (see e.g. \cite{M,TW2})
\bq\label{1K}
K_{N,1}(x,y)=\twotwo{S_{N,1}(x,y)}{(S_{N,1}D)(x,y)}
                   {(\eps S_{N,1})(x,y)-\frac12\sgn(x-y)}{S_{N,1}(y,x)},
   \quad N\textrm{ even},
\eq
and
\bq\label{4K}
  K_{N,4}(x,y)=\frac12\twotwo
     {S_{N,4}(x,y)}{(S_{N,4}D)(x,y)}
    {(\eps S_{N,4})(x,y)}{S_{N,4}(y,x)}.
\eq
Here  $S_{N,\beta}(x,y)$, $\beta=1,4$, are certain
scalar kernels (see \eqref{eq3pp1}, \eqref{eq3pp2} below),
$D$ denotes the differentiation operator, and $\eps$ is
the operator with kernel $\eps(x,y)=\frac12\sgn(x-y)$\footnote{We use the
standard notation $\sgn x=1,$ $0$, $-1$ for $x>0$, $x=0$, $x<0$,
respectively.}.
Such matrix kernels were first introduced by Dyson \cite{Dy70}
in the context of circular ensembles with a
view to computing correlation functions.
Dyson's approach was extended to Hermitian ensembles,
first by Mehta \cite{Me71} for $V(x)=x^2$,
and then for more general weights by Mahoux and Mehta in \cite{MMe91}.
A more direct and unifying approach to the
results of Dyson--Mahoux--Mehta was given
by Tracy and Widom in \cite{TW2},
where formulae \eqref{eq3pp1}, \eqref{eq3pp2} below were derived.
We see that once the kernels
$S_{N,\beta}(x,y)$ are known, then so are the other kernels in $K_{N,\beta}$.
As in the case $\beta=2$, the kernels $K_{N,\beta}$
give rise to explicit formulae for $R_{N,l,\beta}$ and $E_\beta(0;J)$.
For example for $\beta=1,4$
\be
\la{eqR1beta}
        R_{N,1,\beta}(x)\equiv R_{1,\beta}(x) = \frac12\tr K_{N,\beta}(x,x)
\ee
and
$$
        R_{N,2,\beta}(x,y)
         = \frac14\,\big(\tr K_{N,\beta}(x,x)\big)\big(\tr K_{N,\beta}(y,y)\big)
           - \frac12 \tr\big( K_{N,\beta}(x,y) K_{N,\beta}(y,x)\big),
$$
and so on, see \cite{TW2}.
We will discuss some of the literature on edge scaling
after the statement of our results, Theorem \ref{thmuniv}, Corollary \ref{cor1.1}
and \ref{cor1.2} below.
As indicated above, formula \eqref{1K} only applies to the case when $N$
is even. When $N$ is odd, there is a similar, but slightly more complicated,
formula (see \cite{AFNvM}). As in \cite{DG},
throughout this paper, for $\beta=1$,
we will restrict our attention
{\em to the case when $N$ is even.\ }We expect that the methods in this paper
also extend to the case $\beta=1$, $N$ odd, and we plan to consider
this situation in a later publication.
Of course, in situations where the asymptotics of \eqref{1K}
has been analyzed (e.g.{} $V(x)=x^2$) for all $N$ as $N\to\infty$,
the limiting behavior of $R_{N,l,\beta=1}$
is indeed seen to be independent of
the parity of $N$ (see e.g.{} \cite{M,NW}).

Let $\{q_j(x)\}_{j\geq0}$ be any sequence of polynomials
of exact degree $j$, $q_j(x)=q_{j,j}x^j+\cdots$, $q_{j,j}\neq0$.
For $j=0,1,2,\cdots$, set
\be
\la{eq3p1}
        \psi_{j,\beta}(x)=\begin{cases}
                 q_j(x) w_1(x),&\beta=1\cr
                 q_j(x) (w_4(x))^{1/2},&\beta=4.
                                     \end{cases}
\ee
Let $M_{N,1}$ denote the $N\times{}N$ matrix with entries
\be
\la{eq3p2}
       (M_{N,1})_{jk} = (\psi_{j,1},\eps\psi_{k,1}),\qquad 0\leq j,k\leq N-1,
\ee
and let $M_{N,4}$ denote the $2N\times{}2N$ matrix with entries
\be
\la{eq3p3}
       (M_{N,4})_{jk} = (\psi_{j,4},D\psi_{k,4}),\qquad 0\leq j,k\leq 2N-1,
\ee
where again $(\cdot,\cdot)$ denotes the standard real inner product on $\R$.
The matrices $M_{N,1}$ and $M_{N,4}$ are invertible
(see e.g.~\cite[(4.17), (4.20)]{AvM}).
Let $\mu_{N,1}$, $\mu_{N,4}$ denote the inverses of $M_{N,1}$, $M_{N,4}$
respectively.
With these notations we have \cite{TW2} the following formulae
for $S_{N,\beta}$ in \eqref{1K}, \eqref{4K}
\be
\la{eq3pp1}
      S_{N,1}(x,y) = -\sum_{j,k=0}^{N-1}
             \psi_{j,1}(x)\,(\mu_{N,1})_{jk}\,(\eps\psi_{k,1})(y)
\ee
\be
\la{eq3pp2}
      S_{N,4}(x,y) = \sum_{j,k=0}^{2N-1}
             \psi_{j,4}^\prime(x)\,(\mu_{N,4})_{jk}\,\psi_{k,4}(y).
\ee
An essential feature of the above formulae is that the polynomials $\{q_j\}$
are arbitrary
and we are free to choose them conveniently to facilitate the
asymptotic analysis of \eqref{1K}, \eqref{4K} as $N\to\infty$
(see discussion in \cite{DG} and \eqref{eq5p1} below).

In order to state our main result we need more notation.
For any $m\in\N$ let $V(x)$ be a polynomial of degree $2m$
\be
\la{eq3star}
    V(x) = \kappa_{2m}x^{2m}+\cdots,\qquad \kappa_{2m}>0
\ee
and let $w(x)\equiv{}w_{\beta=2}(x)=e^{-V(x)}$ as before.
Let $p_j(x)$, $j\geq0$,
denote the OP's with respect to $w$,
and set $\phi_j(x)\equiv{}p_j(x)(w(x))^{1/2}$, $j\geq0$, as above.
For $\beta=1,4$ set
\be
\la{eq3pppp1}
  V_\beta(x)\equiv \frac12{V(x)}
\ee
and let $N$ be even.
Then by \eqref{eq1p3}, $w_4=e^{-2V_4}=e^{-V}$ and
$w_1=e^{-V_1}=e^{-V/2}$.
This ensures that for the choice $q_j=p_j$ in \eqref{eq3p1}
\be\la{eq5p1}
 \psi_{j,\beta=1}(x)=\psi_{j,\beta=4}(x) = \phi_j(x),
\ee
which enables us in turn to handle $S_{N,1}$ and
$S_{N/2,4}$ in \eqref{eq3pp1}, \eqref{eq3pp2}
simultaneously (see \cite[Remark~1.3]{DG}).
Henceforth and throughout the paper, $K_N$
denotes the Christoffel--Darboux (CD) kernel \eqref{K}, \eqref{eq_KN}
constructed out of these functions $\phi_j$.

For the bulk scaling limit in \cite{DKMVZ} ($\beta=2$) and
 \cite{DG} ($\beta=1,4$),
the authors used the standard scale of one (expected) eigenvalue per unit interval.
At the edge it is standard (see e.g.~\cite{TW3}) to use a slightly different scaling
which ensures that the kernel $\KA(\xi,\eta)$ (see \eqref{eqKA} below)
appears in the limiting forms
 \eqref{equniv2}, \eqref{equniv1}, \eqref{equniv4} below,
without any additional factors. Note that formula \eqref{eqs3p1}
also holds for $\beta=1,4$ and so $R_{N,l=1,\beta}(x)$
gives the density of the expected number of (simple) eigenvalues near
$x$ for $\beta=1,2,4$.
In view of \eqref{eqs3p1}, and also in view of
 \eqref{eqR1beta} and \eqref{1K}, \eqref{4K}
\be
\la{eq14scaling}
\ba
      R_{N,1,2}(x) = K_N(x,x),\quad
        R_{N,1,1}(x) = S_{N,1}(x,x),\quad
        R_{N/2,1,4}(x) = \frac12\,S_{N/2,4}(x,x).
\ea
\ee
To leading order, the right edge of the spectrum is located at
the point $c_N+d_N$ where $c_N,d_N$ are the Mhaskar--Rakhmanov--Saff
numbers in \eqref{cN}, \eqref{dN} below.
For all three cases, in the neighborhood of $c_N+d_N$, we use the
scale
\be
\la{xinetan}
      \xi\mapsto   \xin\equiv c_N\Big(1+\frac\xi{\alpha_N N^{2/3}}\Big)+d_N
\ee
where $\alpha_N$ is given in \eqref{eqI8.1}(2) below. As we will see
(cf.~Remark \ref{rem8-p1} below) this scaling differs slightly
 from a scale of one (expected)
eigenvalue per unit interval.

It turns out that the off-diagonal elements in $K_{N,\beta}$
scale differently as $N\to\infty$.
On the other hand, the statistics of the ensembles are clearly invariant
(cf.~discussion following \eqref{13p.1} below) under the conjugation
$$
    K_{N,\beta}\mapsto K_{N,\beta}^{(\lambda)}\equiv
       \mtt{\lambda^{-1}}{0}{0}{\lambda}\cdot K_{N,\beta} \cdot
      \mtt{\lambda}{0}{0}{\lambda^{-1}}
     =\mtt{(K_{N,\beta})_{11}}{\lambda^{-2}(K_{N,\beta})_{12}}
               {\lambda^{2}(K_{N,\beta})_{21}}{(K_{N,\beta})_{22}}
$$
for any scalar $\lambda$.
For example, this is obviously true for the cluster functions
 $T_{N,l,\beta}$, $\beta=1$ or $4$,
which have the form
\be\la{eqclust}
    T_{N,l,\beta}(y_1,\cdots,y_l)=\frac1{2l}\sum_{\sigma}
       \tr \Big( K_{N,\beta}(y_{\sigma_1},y_{\sigma_2})
                 K_{N,\beta}(y_{\sigma_2},y_{\sigma_3})\cdots
            K_{N,\beta}(y_{\sigma_l},y_{\sigma_1})\Big)
\ee
where the sum is taken over all permutations of $\{1,\cdots,l\}$ (see
\cite[p.~816]{TW2}), etc.

Denote
\be
\la{eqKA}
\ba
     \KA(\xi,\eta)&\equiv\frac{\Ai(\xi)\Aip(\eta)-\Aip(\xi)\Ai(\eta)}{\xi-\eta}\\
               &=\int_0^\infty \Ai(z+\xi)\Ai(z+\eta)\,dz.
\ea
\ee
Set
$$
      \lambda_{(N)}
            \equiv \Big(\frac{c_N}{\alpha_N N^{2/3}}\Big)^{-1/2}.
$$
Theorem \ref{thmuniv}, and Corollary \ref{cor1.1}
and \ref{cor1.2} below are the main results in this paper.
\begin{theorem}
\label{thmuniv}
Let $\beta=2$, $1$ or $4$.
For any $V(x)$ of degree $2m$ as in \eqref{eq3star}
define $V_\beta(x)$ and $w_\beta(x)$ as in \eqref{eq3pppp1}, \eqref{eq1p3}.
Fix a number $L_0$.
Then there exists $c=c(L_0)>0$ such that
as $N\to\infty$\footnote{For $\beta=1,4$, $N$ is even.}
the following holds uniformly for $\xi,\eta\in[L_0,+\infty)$.

In the case $\beta=2$:
\be
\la{equniv2}
\ba
  \mathcal{E}_{N,2}&\equiv
     \frac{1}{\lambda_{(N)}^2}
   K_{N}\big(\xin,\etan\big)
- \KA(\xi,\eta)\to0.
\ea
\eq

In the case $\beta=1$:
\be
\la{equniv1}
\ba
  \mathcal{E}_{N,1}&\equiv
     \frac{1}{\lambda_{(N)}^2}
   K_{N,1}^{(\lambda_{(N)})}
   \big(\xin,\etan\big)
- K^{(1)}(\xi,\eta)\to0
\ea
\ee
where
$$
\ba
  (K^{(1)})_{11}(\xi,\eta)&=(K^{(1)})_{22}(\eta,\xi)
         \equiv \KA(\xi,\eta) + \frac12\Ai(\xi)\cdot\int_{-\infty}^\eta\Ai(t)\,dt\\
  (K^{(1)})_{12}(\xi,\eta)&
         \equiv -\partial_\eta\KA(\xi,\eta) - \frac12\Ai(\xi)\Ai(\eta)\\
  (K^{(1)})_{21}(\xi,\eta)&
         \equiv -\int_\xi^\infty\KA(t,\eta)dt \\
   &\quad- \frac12\int_\xi^\eta \Ai(t)\,dt
            +\frac12\int_\xi^{\infty}\Ai(t)\,dt\cdot\int_\eta^{\infty}\Ai(t)\,dt
         -\frac12\sgn(\xi-\eta).
\ea
$$

In the case $\beta=4$:
\be
\la{equniv4}
\ba
 \mathcal{E}_{N,4}&\equiv
     \frac{1}{\lambda_{(N)}^2}
   K_{N/2,4}^{(\lambda_{(N)})}
   \bigg(\xin,\etan\bigg)
  - K^{(4)}(\xi,\eta)\to0
\ea
\ee
where
$$
\ba
  2(K^{(4)})_{11}(\xi,\eta)&=2(K^{(4)})_{22}(\eta,\xi)
         \equiv \KA(\xi,\eta) - \frac12\Ai(\xi)\cdot\int_\eta^\infty\Ai(t)\,dt\\
  2(K^{(4)})_{12}(\xi,\eta)&
         \equiv -\partial_\eta\KA(\xi,\eta) - \frac12\Ai(\xi)\Ai(\eta)\\
  2(K^{(4)})_{21}(\xi,\eta)&
         \equiv -\int_\xi^\infty\KA(t,\eta)dt
      + \frac12 \int_\xi^{\infty}\Ai(t)\,dt\cdot\int_\eta^{\infty}\Ai(t)\,dt.
\ea
$$
For the error term we have
as $N\to\infty$
\be\la{eqs100}
\ba
 \mathcal{E}_{N,2}&=O(N^{-2/3})e^{-c\xi}e^{-c\eta}\\
 \mathcal{E}_{N,1}
       &= o(1)\twotwo{e^{-c\xi}}{e^{-c\xi}e^{-c\eta}}
                 {1}{e^{-c\eta}}\\
       \mathcal{E}_{N,4}
       &=o(1)e^{-c\xi}e^{-c\eta}
\ea
\ee
uniformly for $\xi,\eta\in[L_0,+\infty)$.
\end{theorem}
\begin{remark}
For $\beta=4$, but not for $\beta=1$,
our methods actually prove that $\mathcal{E}_{N,4}
       =O(N^{-1/(2m)})e^{-c\xi}e^{-c\eta}$.
In order to obtain power law decay for $\mathcal{E}_{N,1}$, it
would be sufficient to obtain power law decay in the error term
in \cite[Theorem~2.2]{DG}: such power law decay can be
obtained using more sophisticated estimates as in \cite{DGKV}.
\end{remark}
We immediately have the following result.
Recall formula \eqref{eqclust} for the cluster functions
for $\beta=1,4$; for $\beta=2$, the cluster functions
have the form \cite[p.~815]{TW2}
$$
    T_{N,l,2}(y_1,\cdots,y_l)=\frac1{l}\sum_{\sigma}
       K_{N}(y_{\sigma_1},y_{\sigma_2})
                 K_{N}(y_{\sigma_2},y_{\sigma_3})\cdots
            K_{N}(y_{\sigma_l},y_{\sigma_1}).
$$
\begin{corollary}\la{cor1.1}
Let $\beta=2$, $1$ or $4$. Let $V$ be a polynomial of degree $2m$
and let $K^{(\beta)}$, $\beta=1,4$
 be as in Theorem~\ref{thmuniv}. Fix a number $L_0$.
Then for $\beta=1$ and $l=2,3,\cdots$
we have uniformly for $\xi_1,\cdots,\xi_l\geq L_0$
\be\la{M.19.1}
\ba
   \lim_{N\to\infty}
       \frac1{(\lambda_{(N)}^2)^l}\,
          &T_{N,l,1}\Big( (\xi_1)^{(N)},
                                                           \cdots,
                   (\xi_1)^{(N)} \Big)\\
    &=\frac1{2l}\sum_{\sigma}
       \tr \Big( K^{(1)}(\xi_{\sigma_1},\xi_{\sigma_2})
                 K^{(1)}(\xi_{\sigma_2},\xi_{\sigma_3})\cdots
            K^{(1)}(\xi_{\sigma_l},\xi_{\sigma_1})\Big).
\ea
\ee
For $\beta=4$, the same result is true provided we replace
$T_{N,l,1}\to{}T_{N/2,l,4}$ and
$K^{(1)}\to{}K^{(4)}$.
For $\beta=2$, the same result is true provided we replace
$T_{N,l,1}\to{}T_{N,l,2}$, $K^{(1)}\to{}\KA$, $\frac1{2l}\to\frac1l$,
and remove the trace.
\end{corollary}

Together with some additional estimates
(see Section \ref{sectthree}),
Theorem~\ref{thmuniv} also yields the following universality
result for the gap probabilities.
Recall that for a $2\times2$ block operator
$A=(A_{ij})_{i,j=1,2}$
with $A_{11},A_{22}$ in trace class and $A_{12},A_{21}$
Hilbert--Schmidt, the regularized $2$-determinant
(see e.g.{} \cite{Simon})
is defined by ${\det}_2(I+A)\equiv\det((I+A)e^{-A})\,e^{\tr(A_{11}+A_{22})}$.

Let $\lambda_1$ denote the largest eigenvalue of a random matrix $M$.
\begin{corollary}\la{cor1.2}
Let $\beta=2$, $1$ or $4$. Let $V$ be a polynomial of degree $2m$
and let $K^{(\beta)}$, $\beta=1,4$
 be as in Theorem~\ref{thmuniv}.
Fix a number $L_0$.
Then the following holds.

In the case $\beta=2$:
\be\la{eqcor1.3beta2}
\ba
   \lim_{N\to\infty}
\Prob\big\{ \lambda_1\leq
(L_0)^{(N)}
\big\}
   ={\det}\big(I-\KA\big|_{L^2([L_0,+\infty))}\big)\equiv F^{(2)}(L_0).
\ea
\ee

In the case $\beta=4$:
\be\la{M.19.4}
\ba
   \lim_{N\to\infty}
  \Prob\big\{ \lambda_1\leq (L_0)^{(N)} \big\}
   =\sqrt{{\det}\big(I-K^{(4)})\big|_{L^2([L_0,+\infty))}\big)}
        \equiv F^{(4)}(L_0).
\ea
\ee

In the case $\beta=1$, let $g(\xi)\equiv\sqrt{1+\xi^2}$,
 $G=\diag(g,g^{-1})$.
Then
\be\la{M.19.4plus}
\ba
   \lim_{N\to\infty} \Prob\big\{\lambda_1\leq
    (L_0)^{(N)} \big\}
   =\sqrt{{\det}_2\big(I-GK^{(1)}G^{-1}\big|_{L^2([L_0,+\infty))}\big)}
                                \equiv F^{(1)}(L_0).
\ea
\ee
\end{corollary}
\begin{remark}
The regularized $2$-determinant is needed for $\beta=1$
because the operator with kernel $\frac12\sgn(\xi-\eta)$
is Hilbert--Schmidt but not trace class in $L^2([L_0,+\infty))$.
The auxiliary function $g$ is needed to ensure that $GK^{(1)}G^{-1}$
indeed has a $2$-determinant: there is considerable freedom
in the choice of the function $g$, see Remark~\ref{remfreedom} below.
\end{remark}
\begin{remark}
\la{rem8-p1}
From Theorem \ref{thmuniv} and \eqref{eq14scaling}
 we have as $N\to\infty$,
\be\la{eq8p-1}
\ba
       \frac{c_N}{\alpha_NN^{2/3}} R_{N,1,2}
            \big(t^{(N)}\big)
         &= \KA(t,t)+o(1)\\
       \frac{c_N}{\alpha_NN^{2/3}}
                R_{N,1,1}\big(t^{(N)}\big) &= \KA(t,t)
                         + \frac12\Ai(t)\int_{-\infty}^t\Ai(u)\,du+o(1) \\
        \frac{c_N}{\alpha_NN^{2/3}}R_{N/2,1,4}\big(t^{(N)}\big) &=
              \frac14 \KA(t,t)
                         - \frac18\Ai(t)\int_t^\infty\Ai(u)\,du+o(1)
\ea
\ee
uniformly for $t$ in any fixed half-line $[L_0,+\infty)$.
In particular the density of the expected number of eigenvalues
at the edge of the spectrum $c_N+d_N$ is given by
\be\la{eq8p-2}
\ba
     \gamma_2 &\equiv (\Ai^\prime(0))^2
                \doteq 0.066987484\\
     \gamma_1 &\equiv (\Ai^\prime(0))^2 +\frac13\Ai(0)
                  \doteq 0.185330168\\
     \gamma_4 &\equiv \frac14(\Ai^\prime(0))^2 - \frac1{24}\Ai(0)
                  \doteq 0.001954035
\ea
\ee
for the indicated values of $\beta=2,1,4$, where we have used
the formula $\KA(t,t)=(\Ai^\prime(t))^2-t(\Ai(t))^2$
and $\int_{-\infty}^0\Ai(u)\,du=\frac23$, $\int_0^\infty\Ai(u)\,du=\frac13$
(see \cite{Stegun}).
Thus setting $t\to\hat{t}/\gamma_\beta$, $\beta=2,1,4$, rescales the axis
so that the density of the expected number of eigenvalues
per unit $\hat{t}$-interval is one.
\end{remark}

The distributions $F^{(\beta)}(L_0)$, $\beta=1,2,4$,
are the celebrated Tracy--Widom distributions which turn out to have applications
in an extraordinary variety of different areas of pure
and applied mathematics (see for example the recent review \cite{TW5}).
The distributions $F^{(\beta)}(L_0)$ can all be expressed
 in terms of a certain solution of the
 Painlev\'e II equation (\cite{TW6,TW4}).

The literature on edge scaling, in particular in the physics
community, is vast, and we make no attempt to present
an exhaustive survey. Rather we will focus on aspects
of the literature which are particularly relevant to this paper.
In the physics literature, early work on edge scaling for $\beta=2$
is due to Moore \cite{Moore} and Bowick and Br\'ezin \cite{BBr}.
In the mathematical literature for $\beta=2$ with Gaussian weight
$V(x)=x^2$, early work can be found in Forrester \cite{F1}
and in the seminal work of Tracy and Widom \cite{TW6},
where the authors derived the Painlev\'e II
representation mentioned above for $F^{(2)}$.
For $\beta=1$ and $4$ in the Gaussian case $V(x)=x^2$,
 the Painlev\'e expressions for $F^{(\beta)}$ were
obtained by Tracy and Widom in \cite{TW4},
but without computing directly the edge scaling limit
of the Fredholm determinants.
The edge scaling limits of matrix kernels $K_{N,\beta}$, $\beta=1,4$,
in the Gaussian case were obtained by Forrester, Nagao and Honner
in \cite{FNH}.
The convergence of the Fredholm determinants
in the Gaussian case for $\beta=1,4$ (and also for $\beta=2$)
was first proved only recently by Tracy and Widom in \cite{TW3}.

Universality at the edge for $\beta=2$ was considered by many authors
in the physics literature (see e.g.~\cite{KaFr}),
and for the cases $\beta=1,4$ see e.g.~\cite{SV}.
The proof of universality at the edge for $\beta=2$ in Theorem~\ref{thmuniv}
above is based on the estimates in \cite{DKMVZ2}
and does not use any results from \cite{W,TW2,DG}. Many
researchers have noted that universality at the edge for $\beta=2$
is true (see e.g.~\cite{CKu}), but we believe
that the details of the proof (Theorem~\ref{thmuniv}, $\beta=2$)
have not been written down previously.
In \cite{St1,St2,St3}, for $\beta=2,1,4$, Stojanovic
proves universality at the edge (and also in the bulk)
in the special case of an even quartic (two-interval) potential
considered previously by Bleher and Its \cite{BI} for $\beta=2$.
Stojanovic uses a variant of the formulae in \cite{W} together
with the asymptotics for OP's obtained in \cite{BI}.
Universality for the distribution of the largest eigenvalue
for a wide class of real and complex Wigner ensembles
(see \cite{M}) was proven by Soshnikov in \cite{So}: the
methods in \cite{So} are completely different from those
in the present paper and are based on the method of moments.
Laguerre ensembles have been considered by many authors,
see e.g.~\cite{F1,FNH}. Various universality issues
at the soft edge, and also at the hard edge and in the bulk,
for generalized Laguerre ensembles for $\beta=2$
were analyzed recently in \cite{Van}.
The authors are currently completing an analysis
of universality questions for such ensembles in the cases $\beta=1$
and $4$, together with Kriecherbauer and Vanlessen, see \cite{DGKV}.

We complete this introduction with a description of Widom's result \cite{W}
which is basic for our approach in this paper.
Widom's method applies to general weights $w_\beta$
with the property that $w_\beta^\prime/w_\beta$ is a rational function.
This property certainly holds for our weights as
in \eqref{eq1p3}, \eqref{eq1p2},
and also for general Laguerre type weights which we consider
in the forthcoming paper \cite{DGKV}.
Introduce  the matrices
\bq\label{DE}
  D_N\equiv{}((D\phi_j,\phi_k))_{0\leq j,k\leq N-1},\qquad
  \eps_N\equiv{}((\eps\phi_j,\phi_k))_{0\leq j,k\leq N-1}.
\eq
It follows from \cite[Section~6]{TW1} that the matrix $D_N$
is banded with bandwidth $2n+1$ where
\be
\la{eqn}
            n\equiv{}2m-1.
\ee
Thus $(D_N)_{jk}=0$ if $|j-k|>n$.
Next, let $N$ be greater than $n$, and
introduce the following  $N$-dependent $n$-column vectors
\bq\label{phi}
\begin{aligned}
   \Phi_1(x)&\equiv{}(\phi_{N-n}(x),\cdots,\phi_{N-1}(x))^T\\
   \Phi_2(x)&\equiv{}(\phi_{N}(x),\cdots,\phi_{N+n-1}(x))^T\\
   \eps\Phi_1(x)&\equiv{}(\eps\phi_{N-n}(x),\cdots,\eps\phi_{N-1}(x))^T\\
   \eps\Phi_2(x)&\equiv{}(\eps\phi_{N}(x),\cdots,\eps\phi_{N+n-1}(x))^T
\ea
\eq
and the following $2n\times2n$ matrices
consisting of four $n\times n$ blocks
\bq\label{B}
  B\equiv{}\twotwo{B_{11}}{B_{12}}
                      {B_{21}}{B_{22}}
   =((\eps\phi_j,\phi_k))_{N-n\leq j,k\leq N+n-1}.
\eq
and
\bq\label{A}
  A\equiv{}\twotwo{0}{A_{12}}
                      {A_{21}}{0}
   =\twotwo{0}{D_{12}}
                      {-D_{21}}{0}
\eq
where
$
  \twotwo{D_{11}}{D_{12}}
                      {D_{21}}{D_{22}}
   \equiv{}((D\phi_j,\phi_k))_{N-n\leq j,k\leq N+n-1}.
$
Finally, set
$$
   C=\twotwo{C_{11}}{C_{12}}
                      {C_{21}}{C_{22}}
     \equiv{}\twotwo{I_n + (BA)_{11}}{(BA)_{12}}
                      {(BA)_{21}}{(BA)_{22}}.
$$
Note that
\be\la{9p.1}
     C_{11}=I_n+B_{12}A_{21}=I_n-B_{12}D_{21}.
\ee
The main result in \cite{W} is the following pair
of formulae for $S_{N,1}$ and $S_{N/2,4}$
\bq
\label{W1}
\ba
   S_{N,1}(x,y) = K_N(x,y)
     - (\Phi_1(x)^T,0^T)&\cdot
        (AC(I_{2n}-BAC)^{-1})^T \\&\cdot(\eps\Phi_1(y)^T,\eps\Phi_2(y)^T)^T
\ea
\eq
and
\bq
\label{W4}
\ba
   S_{N/2,4}(x,y) = K_N(x,y) &+ \Phi_2(x)^T\cdot D_{21}\cdot\eps\Phi_1(y)\\
          &+\Phi_2(x)^T\cdot D_{21}C_{11}^{-1}B_{11}D_{12}\cdot\eps\Phi_2(y).
\ea
\eq
Observe that $S_{N,1}$ and $S_{N/2,4}$ are sums of the $\beta=2$
kernel $K_N(x,y)$ together with correction terms that depend
only on $\phi_{N+j}$ for $j\in\{-n,\cdots,n-1\}$.
The $\beta=4$ case is different from the case $\beta=1$
since, by \eqref{eq3pp2}, for any $x\in\R$,
\be\la{zerobeta4}
      S_{N/2,4}(x,+\infty) = 0, \qquad K_N(x,+\infty)=0.
\ee
Therefore in \eqref{W4} for any (even) $N$ and for all $x\in\R$
\be\la{T-18-00}
   \Phi_2(x)^T\cdot D_{21}\cdot\eps\Phi_1(+\infty)
          +\Phi_2(x)^T\cdot D_{21}C_{11}^{-1}B_{11}D_{12}
            \cdot\eps\Phi_2(+\infty) = 0.
\ee
As the entries of $\Phi_2(x)$ are functionally independent,
and as $D_{12}$ is invertible for large $N$ (see \cite[(2.13)]{DG}),
it follows that
\be\la{eq1.42.p}
   \eps\Phi_1(+\infty) + C_{11}^{-1}B_{11}D_{12} \cdot\eps\Phi_2(+\infty)  = 0
\ee
for large $N$.
From the definition of $\eps$ for any integrable $\psi$
\be\la{T-1-00}
     \eps\psi(y) = \frac12\int_{-\infty}^\infty\psi(t)\,dt
        -\int_{y}^\infty\psi(t)\,dt
      = \eps\psi(+ \infty)  - \int_{y}^\infty\psi(t)\,dt.
\ee
Hence \eqref{W4}, \eqref{T-18-00} imply
\bq
\label{T-1-1}
\ba
   S_{N/2,4}(x,y) = K_N(x,y) &+ \Phi_2(x)^T\cdot D_{21}\cdot
          \bigg(-\int_{y}^\infty\Phi_1(t)\,dt\bigg)\\
          &+\Phi_2(x)^T\cdot D_{21}C_{11}^{-1}B_{11}D_{12}\cdot
           \bigg(-\int_{y}^\infty\Phi_2(t)\,dt\bigg).
\ea
\eq
Formula \eqref{T-1-1} makes clear the decay properties
of $S_{N/2,4}(x,y)$ as $x,y\to+\infty$.
Note that $S_{N,1}$ does not satisfy \eqref{zerobeta4}:
this is the reason why we introduce
auxiliary functions (cf.~$G=\diag(g,g^{-1}))$
 when proving convergence of the determinant in Corollary \ref{cor1.2}.
As noted earlier, the question of convergence of the determinants for $\beta=1,4$
in the Gaussian case was first treated in \cite{TW3}.

The following observations apply to the $21$ entries
in the matrix kernels in the $\beta=1$ and $4$ cases.
Note that by \eqref{eq3pp1},
$(\eps S_{N,1})(x,y)$
is skew symmetric. Thus
\be\la{M.tmp0}
   (\eps S_{N,1})(x,y) = (\eps S_{N,1})(x,y)-(\eps S_{N,1})(y,y)
                        =-\int_{x}^y S_{N,1}(t,y)\,dt.
\ee
Also, from \eqref{eq3pp2}, we see that $(\eps S_{N/2,4})(+\infty,y)=0$
for all $y\in\R$. Together with \eqref{T-1-00}, this implies that
\be\la{1.47beta4}
   (\eps S_{N/2,4})(x,y) =
                     -\int_{x}^\infty S_{N/2,4}(t,y)\,dt.
\ee
These observations simplify evaluation of  integrals of the CD kernel,
and also integrals of the functions $\phi_{N+j}$ in Sections
 \ref{sectfourprime} and \ref{sectfour} below.
\begin{remark}
We note that \eqref{M.tmp0} is also true for $S_{N/2,4}$,
but \eqref{1.47beta4} is more relevant for the calculations that follow.
\end{remark}

In Section \ref{sectthree}, we prove Theorem \ref{thmuniv} and Corollary
\ref{cor1.2} using results on the edge scaling limits of the
CD terms and the correction terms in $K_{N,1}$ and $K_{N,4}$.
These scaling limits are proved in turn in Section \ref{sectfourprime}
for the CD terms, and in Section \ref{sectfour} for the correction terms.
Note that Corollary \ref{cor1.1} is an immediate consequence of Theorem \ref{thmuniv}.

\textbf{Notational remark:} Throughout
 this paper $c,c^\prime,C,C(m),c_1,c_2,\cdots$
refer to constants independent of $N,\xi,\eta$.
The symbols $c,c^\prime,C,\cdots$
 refer to generic constants,
whose precise value may change from one inequality
to another.
The symbol $c_N$ however
always refers to the $N$-dependent constant \eqref{cN} below.

{\bf Acknowledgments.}
The work of the first author was supported in part by
NSF grants DMS--0296084 and DMS--0500923.
The second author would like to thank
the Courant Institute,
New York University, where he has spent a part
of the academic year 2004--05,
 for hospitality and financial support.
The second author also would like to thank Caltech for hospitality
and financial support. Finally, the second author would like
to thank the Swedish foundation STINT for providing basic support to visit Caltech.
\section{Proofs of Theorem \ref{thmuniv} and Corollary \ref{cor1.2}}
\la{sectthree}
The key estimates for the proofs of Theorem \ref{thmuniv} and Corollary \ref{cor1.2} are obtained below in Section \ref{sectfourprime} for the CD terms and
in Section \ref{sectfour} for the correction terms.
\subsection{Proof of Theorem \ref{thmuniv}}
Inequality \eqref{goalCDandderivs} proves the result for the $\beta=2$ case.

In the case $\beta=4$, we use \eqref{T-1-1}
and consider the CD part and the correction term separately.
The properly scaled
 $11$, $22$ and $12$ entries of $K_{N/2,4}^{(\lambda_{(N)})}(\xin,\etan)$
converge to the corresponding entries in \eqref{equniv4} et seq.~with the error
 estimate $o(1)e^{-c\xi}e^{-c\eta}$,
uniformly for $\xi,\eta\in [L_0,+\infty)$:
    this follows from \eqref{goalCDandderivs}
for the CD kernel part,
and from \eqref{T-8-2l} and \eqref{T-7-0}, respectively, for the correction term.
By \eqref{1.47beta4}, \eqref{D-13-0} and \eqref{T-11-1},
 the (unscaled) $21$ entry $(\eps S_{N/2,4})(\xin,\etan)$
of $K_{N/2,4}$ satisfies
\be\la{eq11.p}
\ba
   \bigg|2(\eps S_{N/2,4})(\xin,\etan)
   &- \Big[\Big(-\int_{\xi}^\infty \KA(t,\eta)\,dt \\
 &+\frac12\big(\int_{\xi}^\infty \Ai(t)\,dt\big)
       \big(\int_{\eta}^\infty\Ai(t)\,dt\big)\Big]\bigg| \leq o(1)e^{-c\xi}e^{-c\eta}
\ea
\ee
uniformly for $\xi,\eta\in[L_0,+\infty)$.
This completes the proof of Theorem~\ref{thmuniv} for $\beta=4$.

In the case $\beta=1$, we use \eqref{W1}
and again consider the CD part and the correction term separately.
The properly scaled
 $11$ and $22$ entries of $K_{N,1}^{(\lambda_{(N)})}(\xin,\etan)$
converge to the corresponding entries in \eqref{equniv1} et seq.~with the error
estimates $o(1)e^{-c\xi}$
and $o(1)e^{-c\eta}$, respectively,
uniformly for $\xi,\eta\in [L_0,+\infty)$:
this follows from \eqref{goalCDandderivs} for the CD kernel part
(giving rise to a smaller error $o(1)e^{-c\xi}e^{-c\eta}$)
and from \eqref{corr11beta1} for the correction term.
The properly scaled $12$ entry converges to the
corresponding entry in \eqref{equniv1} et seq.~with
 error $o(1)e^{-c\xi}e^{-c\eta}$,
uniformly for $\xi,\eta\in [L_0,+\infty)$: this
 follows from \eqref{goalCDandderivs} for the CD kernel part
and from \eqref{T-12-2} for the correction term.
Finally, in view of \eqref{M.tmp0},
 \eqref{D-13-0} and \eqref{T-25-0},
the (unscaled) $21$ entry of $K_{N,1}^{(\lambda_{(N)})}(\xin,\etan)$
satisfies
\be\la{corrl21beta1}
\ba
  \bigg| (\eps S_{N,1})(&\xin,\etan)
   -\bigg[
 -\int_{\xi}^\eta \KA(t\eta)\,dt
    -      \frac12 \int_\xi^\eta\Ai(s)\,ds \\
   &+   \frac12 \bigg(\int_\xi^\eta \Ai(s)\,ds\bigg)
                     \bigg( \int_\eta^\infty \Ai(t)\,dt\bigg) \bigg]\bigg|
    \leq o(1)e^{-c\min(\xi,\eta)} = o(1)
\ea
\ee
with the uniform estimate $o(1)$ for $\xi,\eta \geq L_0$.
In order to obtain the same form for the limit as
claimed in Theorem~\ref{thmuniv},
we note that for all $\xi,\eta\in\R$
\be\la{eqprop10}
\ba
-\int_{\xi}^\eta &\KA(t,\eta)\,dt
 +\frac12\big(\int_{\xi}^\eta\Ai(t)\,dt\big)\big(\int_{\eta}^\infty\Ai(t)\,dt\big)\\
   &=- \int_{\xi}^\infty \KA(t,\eta)\,dt
+\frac12\big(\int_{\xi}^\infty\Ai(t)\,dt\big)\big(\int_{\eta}^\infty\Ai(t)\,dt\big).
\ea
\ee
Indeed, a direct calculation using the representation \eqref{eqKA}
for $\KA$ shows that the RHS of \eqref{eqprop10} is skew symmetric
in $\xi$ and $\eta$.
In particular, the RHS vanishes for $\xi=\eta$,
as is also evident for the LHS.
But the $\xi$ derivatives of both sides are equal
and hence the identity follows.
This finishes the proof of Theorem~\ref{thmuniv}.
\subsection{Proof of Corollary \ref{cor1.2}}
The following basic fact is well-known (see e.g.~\cite{ReSi}).
Let $D=d/dx$ denote differentiation and let $\rho(x)$
be any positive function such that $\rho^{-1}\in L^2(\R)$.
Then the operator
\be\la{HSestaux}
        A = \frac 1\rho \frac 1{D+I}
\ee
is Hilbert--Schmidt in $L^2(\R)$. Indeed, by the Fourier transform,
$A$ is unitarily equivalent to an operator with square integrable kernel
$\widehat{(\rho^{-1})}(k-k^\prime)\frac1{ik^\prime+1}$, $k,k^\prime\in\R$.
\subsubsection{The case $\beta=2$}
\la{ssec2.2.1}
Let $\lambda_1$ denote the largest eigenvalue of the matrix $M$
in the unitary ensemble.
It is well-known (see e.g.~\cite{TW2}) that for finite $N$
$$
\ba
\Prob\Big\{ &\lambda_1\leq c_N\Big(1+\frac{L_0}{\alpha_NN^{2/3}}\Big)
                   +
                     d_N \Big\}\\
       &= \det\Big(1- \frac{c_N}{\alpha_NN^{2/3}}
                  K_N(\xi^{(N)},\eta^{(N)})\Big|_{L^2([L_0,+\infty))}\Big).
\ea
$$
Since $K_N$ is finite rank, it is indeed
trace class.
As the trace class
determinant is continuous under the trace class convergence,
we only have to prove that
\be\la{eq2.7minus}
   \Delta_N(\xi,\eta)\equiv    \frac{c_N}{\alpha_NN^{2/3}}
                  K_N(\xi^{(N)},\eta^{(N)})
      -\KA(\xi,\eta)\to0,\qquad \textrm{as }N\to\infty,
\ee
in the trace norm in $L^2([L_0,+\infty))$,
in order to prove Corollary \ref{cor1.2} for $\beta=2$.
Let $\chi_{L_0}^{\#}(\xi)$ be a $C^\infty$ function
such that $\chi_{L_0}^{\#}(\xi)=1$ for $\xi\geq L_0$
and $\chi_{L_0}^{\#}(\xi)=0$ for $\xi\leq L_0-1$.
We will show that
\be\la{eq12p.1}
    \chi_{L_0}^{\#} \Delta_N \chi_{L_0}^{\#}
                   \to0,\qquad N\to\infty,
\ee
in the trace norm in $L^2(\R)$.
But then $\chi_{L_0} \Delta_N \chi_{L_0}
  =  \chi_{L_0}\big(\chi_{L_0}^{\#} \Delta_N \chi_{L_0}^{\#}\big)
 \chi_{L_0}$
also converges to zero in trace norm in $L^2(\R)$, where
$\chi_{L_0}$ is the characteristic function of $[L_0,+\infty)$,
and this clearly proves \eqref{eq2.7minus}.

Let $\rho(\xi)=(1+\xi^2)^{1/2}$ and write
$$
    \chi_{L_0}^{\#} \Delta_N \chi_{L_0}^{\#}
     =    \Big[\frac1\rho \frac1{D+I}\Big]
       \, \big[(D+I)\rho \chi_{L_0}^{\#} \Delta_N \chi_{L_0}^{\#} \big].
$$
The first operator is Hilbert--Schmidt (see \eqref{HSestaux})
and the second operator is of order $O(N^{-2/3})$
 in Hilbert--Schmidt norm by \eqref{goalCDandderivs},
with $L_0$ replaced with $L_0-1$.
This proves \eqref{eq12p.1}.
\subsubsection{The case $\beta=4$}
Let $\lambda_1$ denote the largest eigenvalue of the matrix $M$
in the symplectic ensemble.
Then in \cite{TW2} the authors prove
$$
\ba
\Prob\Big\{ &\lambda_1\leq c_N\Big(1+\frac{L_0}{\alpha_NN^{2/3}}\Big)
                   +
                     d_N \Big\}\\
       &= \sqrt{\det\Big(1- \frac{c_N}{\alpha_NN^{2/3}}
                  K_{N/2,4}^{(\lambda_{(N)})}
                     (\xi^{(N)},\eta^{(N)})\Big|_{L^2([L_0,+\infty))}\Big)}.
\ea
$$
The proof will therefore be complete if we could prove that
all the four entries of $K_{N/2,4}^{(\lambda_{(N)})}(\xin,\etan)$ converge
to the corresponding entries of $K^{(4)}(\xi,\eta)$ in trace
class norm in $L^2([L_0,\infty))$.
Again we use \eqref{T-1-1}
and prove the trace class convergence of the CD part and
  of the correction term separately.
The trace class convergence of the CD parts of all the four entries
of $K_{N/2,4}^{(\lambda_{(N)})}$ follows by using \eqref{goalCDandderivs}
and \eqref{D-13-0} together with the trace class convergence method in
Subsection \ref{ssec2.2.1}.

To prove the convergence in trace class
for the $11$ and $22$ correction terms, we must show
that
$$
\ba
   \Delta_N(\xi,\eta) \equiv \frac{c_N}{\alpha_NN^{2/3}}
   \bigg[  \Phi_2&(\xin)^T\cdot D_{21}\cdot
          \bigg(-\int_{\etan}^\infty\Phi_1(t)\,dt\bigg)\\
          \null&\qquad+\Phi_2(\xin)^T\cdot
                           D_{21}C_{11}^{-1}B_{11}D_{12}\cdot
           \bigg(-\int_{\etan}^\infty\Phi_2(t)\,dt\bigg)\\
   &- \bigg(- \frac12 \Ai(\xi)\int_\eta^\infty\Ai(t)\,dt\bigg)\bigg]
\ea
$$
(cf.~\eqref{eq4.10p}, \eqref{T-8-2l})
converges to zero in  trace class
in $L^2([L_0,\infty))$.
But $\Delta_N$ is an operator with  finite rank at most $n+1=2m=\deg V$,
independent of $N$.
For such operators we have the following inequality
\be\la{12pp.1}
   \|\Delta_N\|_1 \leq \sqrt{2m}\|\Delta_N\|_{HS}
\ee
where $\|\cdot\|_{1}$, $\|\cdot\|_{HS}$ denote the trace norm, Hilbert--Schmidt norm
in $L^2([L_0,\infty))$,
respectively. Indeed, $|\Delta_N|=\sqrt{\Delta_N^*\Delta_N}$
is also an operator of rank at most $2m$, and hence it has at most
$2m$ nonzero eigenvalues, $\sigma_{1}\geq\sigma_2\geq\cdots
\geq \sigma_j>0$, $0\leq j\leq 2m$.
Thus
$$
\ba
      \|\Delta_N\|_1 =\tr |\Delta_N| = \sum_{i=1}^j \sigma_i
            \leq \sqrt{j} \bigg(\sum_{i=1}^j \sigma_i^2\bigg)^{1/2}
              \leq \sqrt{2m}\|\Delta_N\|_{HS}.
\ea
$$
But from \eqref{T-8-2l},
  $\|\Delta_N\|_{HS}=o(1)\big(\int_{L_0}^\infty\int_{L_0}^\infty
e^{-c\xi}e^{-c\eta}\,d\xi d\eta\big)^{1/2}=o(1)$, $N\to\infty$,
and we conclude that $\|\Delta_N\|_1\to0$, $N\to\infty$,
as desired. A similar argument using \eqref{T-7-0} for the $12$ entry
and \eqref{T-11-1} for the $21$ entry,
 completes the proof of Corollary~\ref{cor1.2}
for $\beta=4$.
\subsubsection{The case $\beta=1$}
Let $\lambda_1$ denote the largest eigenvalue of the matrix $M$
in the orthogonal ensemble.
Let $g(\xi)=\sqrt{1+\xi^2}$ and
  set $G(\xi)=\twotwo{g(\xi)}{0}{0}{g^{-1}(\xi)}$.
Note that $g^{-1}(\xi)\in L^2(\R)$.
Let $g_{(N)}(t)=\sqrt{1+[\frac{\alpha_NN^{2/3}}{c_N}(t-c_N-d_N)]^2}$
and $G_{(N)}(\xi)=\twotwo{g_{(N)}(\xi)}{0}{0}{g_{(N)}^{-1}(\xi)}$.
   Note that $g_{(N)}(\xin)=g(\xi)$. Recall the
 definition of $\det_2$ in the Introduction.
A slight modification of the calculations in \cite[Section~9]{TW2}
shows that
\be\la{13p.1}
\ba
\Prob\Big\{ &\lambda_1\leq c_N\Big(1+\frac{L_0}{\alpha_NN^{2/3}}\Big)
                   +
                     d_N \Big\}\\
       &= \sqrt{{\det}_2\Big(1- \frac{c_N}{\alpha_NN^{2/3}}
                  \Big(G_{(N)} K_{N/2,4}^{(\lambda_{(N)})} G_{(N)}^{-1}\Big)
                     (\xi^{(N)},\eta^{(N)})\Big|_{L^2([L_0,+\infty))}\Big)}.
\ea
\ee
In \cite[Section~9]{TW2} the authors use the fact that
$\det(1+AB)=\det(1+BA)$ for appropriate operators $A$ and $B$.
But one clearly has the freedom to write $AB=AG_{(N)}^{-1}G_{(N)}B$,
and so we also have $\det(1+AB)=\det(1+AG_{(N)}^{-1}G_{(N)}B)=
 \det(1+G_{(N)}BAG_{(N)}^{-1})$ and this leads to \eqref{13p.1}.
We have chosen $G_{(N)}$ as above in such a way as to ensure
 that $1+G_{(N)}BAG_{(N)}^{-1}$
has a $2$-determinant, but there is clearly great freedom in
the choice of $g_{(N)}$, and hence of $G_{(N)}$.
From \eqref{13p.1} we see that
in order to prove \eqref{M.19.4plus} it is enough to show \cite{Simon}
that the diagonal (respectively the off-diagonal) entries
of $\frac{c_N}{\alpha_NN^{2/3}}
                  \big(G_{(N)} K_{N/2,4}^{(\lambda_{(N)})} G_{(N)}^{-1}\big)
                     (\xi^{(N)},\eta^{(N)})$
converge to the respective entries
 of $(GK^{(1)}G^{-1}) (\xi,\eta)$ in trace (respectively Hilbert--Schmidt)
norm in $L^2([L_0,\infty))$.

We consider first the $11$ entry (again the $22$ entry can be considered
similarly).
This entry has the form
$$
\ba
  \frac{c_N}{\alpha_NN^{2/3}}
                  g_{(N)}(\xin)  &S_{N,1}(\xin,\etan) g_{(N)}^{-1}(\etan)
= \frac{c_N}{\alpha_NN^{2/3}}
                  g(\xi)  S_{N,1}(\xin,\etan) g^{-1}(\eta)
\ea
$$
where $S_{N,1}$ is given by the CD part and the correction term
as in \eqref{W1}.
The proof that $g(\xi)  \big[\frac{c_N}{\alpha_NN^{2/3}}
                  K_N(\xin,\etan) - \KA(\xi,\eta) \big] g^{-1}(\eta)\to0$,
$N\to\infty$, in trace norm in $L^2([L_0,\infty))$
is completely analogous to the $\beta=2$ case in Subsection~\ref{ssec2.2.1}
(note that $g$ and its derivative are polynomially bounded)
and the details are left to the reader.

As in the $\beta=4$ case above, the fact that the correction
term in the $11$ entry has a fixed maximal rank independent of $N$
implies that the trace norm convergence follows from the Hilbert--Schmidt
convergence. But by \eqref{T-13-1}, \eqref{corr11beta1}
$$
\ba
  \bigg|g(\xi)\bigg[
   -\Phi_1(\xin)^T\cdot &G_{11}\cdot
                      \bigg(-\int_{\etan}^\infty\Phi_1(t)\,dt\bigg)
          -\Phi_1(\xin)^T\cdot G_{12}
                       \cdot \bigg(-\int_{\etan}^\infty\Phi_2(t)\,dt\bigg)\\
   &-\Phi_1(\xin)^T\cdot G_{11}\cdot\eps\Phi_1(+\infty)
          -\Phi_1(\xin)^T\cdot G_{12} \cdot\eps\Phi_2(+\infty)\\
   &- \frac12 \Ai(\xi) \int_{-\infty}^\eta\Ai(t)\,dt \bigg]g^{-1}(\eta)\bigg|
        \leq o(1)g(\xi)e^{-c\xi}g^{-1}(\eta)
\ea
$$
which is $o(1)$ in Hilbert--Schmidt norm in $L^2([L_0,\infty))$.
This proves the trace class convergence of the $11$ (and similarly of the $22$)
entry.

Finally, we note from the uniform pointwise bounds in \eqref{eqs100}
that the error terms in the $12$ and $21$ entries
are bounded by $o(1)g(\xi)e^{-c\xi}e^{-c\eta}g(\eta)$
and $o(1)g^{-1}(\xi)g^{-1}(\eta)$, respectively,
uniformly for $\xi,\eta\geq L_0$.
This immediately implies the Hilbert--Schmidt convergence
of the off-diagonal entries to their appropriate limits.
This completes the proof of Corollary \ref{cor1.2}.
\begin{remark}
With a little more work one can show that
in the $\beta=1$ case the off-diagonal
entries (apart from the term $g^{-1}(\xi)\sgn(\xi-\eta)g^{-1}(\eta)$)
 in fact converge in trace class norm,
and not just in Hilbert--Schmidt norm.
\end{remark}
\begin{remark}
\la{remfreedom}
As noted earlier, there is considerable freedom in the choice
of the auxiliary function $g$. We see that all we need is that $g,g^\prime$
are polynomially bounded and $g^{-1}\in L^2(\R)$.
\end{remark}
\section{The edge scaling limits of the Christoffel--Darboux ($\beta=2$)
kernel, and of its derivatives and integrals}
\la{sectfourprime}
\subsection{Auxiliary facts from \cite{DKMVZ2}}
We now recall some notation from [ibid.].
Let $d\mu_N^{(\textrm{eq})}(x)$ denote the equilibrium measure
(see e.g.~\cite{SaTo}) for OP's
corresponding to the rescaled weight $e^{-NV_N(x)}$,
$V_N=\frac1NV(c_Nx+d_N)$,
where $c_N$, $d_N$ are the so-called Mhaskar--Rakhmanov--Saff
(MRS) numbers (see \cite{MhSa,Ra}).
For $V(x)=\kappa_{2m}x^{2m}+\kappa_{2m-1}x^{2m-1}+\cdots$
as in \eqref{eq3star}, we have [ibid., Thm.~2.1]
to any order $q$ as $N\to\infty$
\be\la{cN}
   c_N = \bigg(\frac1{\kappa_{2m}}
       \frac{(2m)!!}{m(2m-1)!!}\bigg)^{1/(2m)}\,N^{1/(2m)}
             + \sum_{j=0}^q c_{(j)} N^{-j/(2m)} + O(N^{-(q+1)/(2m)})
\ee
and
\be\la{dN}
   d_N = -\frac{\kappa_{2m-1}}{2m\kappa_{2m}}
             + \sum_{j=1}^q d_{(j)} N^{-j/(2m)} + O(N^{-(q+1)/(2m)}).
\ee
As $N\ra\infty$, the equilibribum measure is absolutely continuous
with respect to Lebesgue measure, $d\mu_N^{(\textrm{eq})}(x)
=\psi_N^{(\textrm{eq})}(x)\,dx$, and is supported on the (single)
interval $[-1,1]$,
\be
\la{psiN}
      \psi_N^{(\textrm{eq})}(x)\equiv\psi_N(x)
                     =\frac1{2\pi}|1-x^2|^{1/2}\chi_{[-1,1]}(x)
          \,h_N(x)
\ee
(see [ibid., (2.4)])
 where $h_N(x)$ is a real polynomial of degree $2m-2$ satisfying
 [ibid., Prop.~5.3]
\be\la{hestbelow}
      h_N(x)\geq h_{\textrm{min}}>0, \qquad x\in\R,\qquad N\geq N_1(V).
\ee
Set [ibid., (5.33)]
$$
      g(z)\equiv g_N(z)=\int_{-1}^1 \log(z-x)\,d\mu_N^{\textrm{(eq)}}(x)
                =\int_{-1}^1 \log(z-x)\,\frac1{2\pi}|1-x^2|^{1/2}\,h_N(x)\,dx,
$$
$z\in\C\setminus(-\infty,-1]$, and for $z\in(-1,1)$  [ibid., (5.34)]
\be\la{M.2.2}
   \Xi_N(z) \equiv g_+(z)-g_-(z) = i\int_z^1 |1-x^2|^{1/2} h_N(x)\,dx.
\ee
We also use the same symbol for the analytic continuation
of $\Xi_N$ to $\C\setminus((-\infty,-1]\cup[1,+\infty))$.

\noindent
{\bf Notational remark:} Here we denote by $\Xi_N$
what was denoted by $\xi_N$ in [ibid.].

For a fixed $\delta>0$ sufficiently small
(cf.~\cite[Rem.~4.3]{DG}), let $R$ denote the matrix function
defined in \cite[(7.47)]{DKMVZ2}.
The function $R$ is analytic in the complement
of the contour $\hat{\Sigma}_R$ as in [ibid., Fig.~7.6] and is continuous
up to the boundary. Furthermore by [ibid., Thm.~7.10], it has an asymptotic
expansion
\be\la{resolv} 
    R(z) \sim I + N^{-1}\sum_{k=0}^\infty r_k(z)N^{-k/(2m)}
\ee
where $\{r_k(z)\}$ are bounded functions that are analytic in the complement
of $\{|z-1|=\delta\}\cup\{|z+1|=\delta\}$. The expansion \eqref{resolv} is
 uniform for $z\in\C\setminus\hat{\Sigma}_R$.
Moreover, by the proof of [ibid., Thm~7.10] and Cauchy's theorem,
it follows that \eqref{resolv} can be differentiated term by term,
\be\la{eq29pp.1}
   \frac{d^j}{dz^j} R(z) \sim N^{-1}\sum_{k=0}^\infty \frac{d^j}{dz^j}
            r_k(z)N^{-k/(2m)},\qquad j=1,2,\cdots,
\ee
where again the expansion is uniform for $z\in\C\setminus\hat{\Sigma}_R$.
Also, each $\frac{d^j}{dz^j} r_k(z)$
 is bounded (and analytic) in the complement
of $\{|z-1|=\delta\}\cup\{|z+1|=\delta\}$.
\subsection{Estimates on the CD kernel and its derivatives}
\la{subsecCDderivs}
We will only consider the end point $1$ (the end point $-1$
can be considered similarly).
Let $L_0\in\R$ be fixed.
Recall the notation in \eqref{eqKA}, \eqref{xinetan} and \eqref{K}.
Our goal in this Subsection is to prove
that for $j,k=0,1$, and some $C=C(L_0)>0$, $c=c(L_0)>0$,
one has uniformly for $\xi,\eta\in[L_0,+\infty)$
\be
\la{goalCDandderivs}
   \bigg|\partial_\xi^j\partial_\eta^k\Big[\frac{c_N}{\alpha_N N^{2/3}}
       K_N(\xin,\etan)
          -\KA(\xi,\eta)\Big]\bigg| \leq CN^{-2/3}e^{-c\xi}e^{-c\eta}.
\ee
Note that \cite{Stegun}
\be
\la{C-3-1}
\ba
         |&\Ai(\xi)| \leq C (1+|\xi|)^{-1/4},\qquad
              |\Ai^\prime(\xi)| \leq C (1+|\xi|)^{1/4},\qquad \xi\in\R,\\
   |&d^q\Ai(\xi)/d\xi^q| \leq C_1 e^{-\xi}\leq C_2,\qquad \xi\in[L_0,+\infty),
         \qquad q=0,1,2,\cdots.
 \ea
\ee
\subsubsection{Auxiliary notation}
Set (see \cite[(2.15)]{DKMVZ2} and also \cite[(4.10)]{DG})
\be\la{eqI8.1}
        f_N(x) = \alpha_NN^{2/3}\,(x-1)\fh_N(x)
\ee
which satisfies the following (see (the proof of) \cite[Proposition~7.3]{DKMVZ2})
\begin{enumerate}
\item
 $\fh_N(x)$ is real analytic on $(1-2\delta,1+2\delta)$,
and to any order $q=0,1,2,\cdots$
$$
   \fh_N(x) = \sum_{j=0}^q N^{-j/(2m)}\,\fh_{(j)}(x) + O(N^{-(q+1)/(2m)})
$$
uniformly for $x$ in the interval. Moreover, the functions $\fh_{(j)}(x)$
are also real analytic on $1-2\delta< x< 1+2\delta$
\item
to any order $q=1,2,\cdots$
$$
         \alpha_N\equiv \big(h_N^2(1)/2\big)^{1/3}
            =2m^{2/3} +  \sum_{j=1}^q N^{-j/(2m)}\,\alpha_{(j)}
                   + O(N^{-(q+1)/(2m)})
$$
\item
 $f_N^\prime(x)=-\alpha_N N^{2/3}W_N(x)$,
where $W_N(x)=\fh_N(x) + (x-1)\fh_N^\prime(x)$ also has
an expansion uniform in $x$ to any order $q=0,1,2,\cdots$ as above
$$
  W_N(x) = \sum_{j=0}^q N^{-j/(2m)}\,W_{(j)}(x) + O(N^{-(q+1)/(2m)}).
$$
The terms $W_{(j)}(x)$ are real analytic on $1-2\delta< x< 1+2\delta$
\item
 $\max_{k=0,1,2}\max_{1-2\delta\leq x\leq1+2\delta}
  |d^k\fh_N(x)/dx^k|\leq M<\infty$ for $N\geq N_2(V)$
\item
 $\fh_N(1)=1=W_N(1)$
 and $\min_{1-2\delta\leq x\leq1+2\delta}\fh_N(x)\geq\frac12$
for $N\geq N_2(V)$.
Also $\fh_{(0)}(1)=1=W_{(0)}(1)$.
\end{enumerate}
Denote
\be\la{xenn}
\ba
   \xi_N&\equiv \xi/(\alpha_N N^{2/3}),
             \qquad \quad \eta_N\equiv \eta/(\alpha_N N^{2/3})\\
   I_N&\equiv[L_0,\alpha_N N^{2/3}\delta],
     \qquad II_N\equiv[\alpha_N N^{2/3}\delta,+\infty).
\ea
\ee
Thus, recalling \eqref{xinetan},
 $\xi^{(N)}=c_N(1+\xi_N)+d_N$
 and similarly $\eta^{(N)}=c_N(1+\eta_N)+d_N$.
As above, let $\delta>0$ be fixed and sufficiently small.
Consider first $\xi_N,\eta_N$ in a neighborhood of $0$.
Set
\be
\la{C-1-0}
\ba
 g_N(\xi)&\equiv \xi \fh_N\big(1+\xi_N\big)\\
  \Fh_N\big(1+\xi_N\big)
  &\equiv    \big(2+\xi_N\big)^{1/4}\cdot
           \big(\fh_N\big(1+\xi_N\big)\big)^{1/4}\\
  F_N\big(1+\xi_N\big)
     &\equiv N^{1/6}\alpha_N^{1/4}
        \Fh_N\big(1+\xi_N\big)
\ea
\ee
and also
\be\la{eq5.4p}
\ba
      A_0(\xi)&\equiv N^{1/6}\alpha_N^{1/4}
        \Fh_N\big(1+\xi_N\big)\cdot \Ai(g_N(\xi))\\
      A_1(\xi)&\equiv N^{-1/6}\alpha_N^{-1/4}
       \Big( \Fh_N\big(1+\xi_N\big)\Big)^{-1}\cdot \Aip(g_N(\xi)).
\ea
\ee
Note that in view of \eqref{eqI8.1}(1)(5) and the formula
\be\la{gNderiv}
\ba
   g_N^\prime(\xi)=\fh_N(1+\xi_N)
       + \xi_N \,\fh_N^\prime(1+\xi_N)
\ea
\ee
there exist $c_2>c_1>0$ such that
\be
\la{gNprop1}
\ba
     c_1\leq\frac{g_N(\xi)}{\xi}\leq c_2,\qquad   \xi\in I_N
\ea
\ee
and
\be
\la{gNprop}
   c_1\leq g_N^\prime(\xi) \leq  c_2, \qquad
      |g_N^{\prime\prime}(\xi)| \leq  c_2 N^{-2/3}, \qquad
        \xi\in I_N.
\ee
Similarly
one has uniformly for $\xi\in I_N$
\be
\la{FNprop}
  c_1 \leq \Fh_N\big(1+\xi_N\big)
              \leq c_2,\qquad
   \bigg| \frac{d^k}{dz^k}\Fh_N(z)\big\vert_{z=1+\xi_N}
         \bigg| \leq C(k)
\ee
for some $C(k)$, $k=1,2,\cdots$.
\subsubsection{Estimates for
$(\xi,\eta)\in I_N\times I_N$}
\la{ssecAiAi}
With the above notation the following holds.
\begin{proposition}
\la{AiAi}
For $(\xi,\eta)\in I_N\times I_N$
\be\la{D-11-1}
        \frac{c_N}{\alpha_N N^{2/3}}
           K_N(\xin,\etan) = \KA(\xi,\eta) + \frac1{\alpha_N N^{2/3}}
                   \sum_{j=1}^4 Q_{1,j}(\xi,\eta)
\ee
where
\be\la{QQ2}
\ba
   Q_{1,1}(\xi,\eta)&\equiv -
             \onetwo{A_0(\eta)}{A_1(\eta)}\cdot
             \twotwo{1}{-i}{-1}{-i}\\
          &\quad\cdot
     \int_0^1 (R^T)^\prime\big(1+\xi_N
             +t(\eta_N-\xi_N)\big)\,dt\\
          &\quad
         \cdot{\twotwo{1}{-i}{-1}{-i}}^{-1}
         \twoone{-A_1(\xi)}{A_0(\xi)}
\ea
\ee
and
\be\la{Q12}
\ba
     Q_{1,2}(\xi,\eta)&\equiv   \Ai(g_N(\xi)) \Aip (g_N(\eta))\cdot
             T_N(\xi,\eta) -  \Ai(g_N(\eta)) \Aip (g_N(\xi))\cdot
             T_N(\eta,\xi) \\
     T_N(\xi,\eta)&\equiv \frac{\int_0^1 \Fh_N^\prime(1+\eta_N
                      +  \tau(\xi_N-\eta_N))\,d\tau}
               {\Fh_N(1+\eta_N)}
\ea
\ee
and
\be\la{Q13}
\ba
     Q_{1,3}(\xi,\eta)&\equiv   E_N(\xi,\eta)
          \int_0^\infty\Ai(z+g_N(\xi)) \Ai (z+g_N(\eta))\,dz\\
       E_N(\xi,\eta)&\equiv \int_0^1
          \big[\eta  +  \tau(\xi-\eta)\big]
      \Big[  \fh_N^\prime(1 + \eta_N   +  \tau(\xi_N-\eta_N)) \\
             &\qquad\qquad+ \int_0^1 \fh_N^\prime\big(1+\sigma(\eta_N
                      +  \tau(\xi_N-\eta_N))\big)\,d\sigma\Big]\,d\tau
\ea
\ee
and
\be\la{Q14}
\ba
     Q_{1,4}(\xi,\eta)&\equiv  \xi^2 L_N(\xi)
          \int_0^\infty U_N(\xi,z) \Ai (z+g_N(\eta))\,dz\\
        &\qquad+ \eta^2 L_N(\eta)
          \int_0^\infty \Ai(z+\xi) U_N(\eta,z) \,dz\\
       L_N(\xi)&\equiv  \int_0^1 \fh_N^\prime(1+\sigma \xi_N )\,d\sigma\\
      U_N(\xi,z) &\equiv  \int_0^1 \Aip \big( z+ \xi +\tau(g_N(\xi)-\xi)\big)\,d\tau.
\ea
\ee
\end{proposition}
\begin{proof}
First, some algebra: let $Y$ solve
 the Fokas--Its--Kitaev Riemann--Hilbert problem
for the polynomials orthogonal with respect to the weight $e^{-V(x)}dx$
(see \cite[Thm.~3.1]{DKMVZ2}). Then as in \cite[(6.3)]{DKMVZ}
we find for any $x,y\in\R$
\be\la{A-1-1}
\ba
   K_N(x,y) &= e^{-(V(x)+V(y))/2}\,
       \frac{Y_{11}(y)Y_{21}(x) - Y_{11}(x)Y_{21}(y)}{2\pi i(x-y)}\\
      &= -e^{-(V(x)+V(y))/2}\,
   \frac{\onetwo{1}{0}\cdot Y_+^T(y)\cdot Y_+^{-T}(x)
                  \cdot{\onetwo{0}{1}}^T}{2\pi i(x-y)}.
\ea
\ee
Here and below $+/-$ refer
 to the boundary values taken from above/below the real axis.
(The choice $Y_+$ is made only for definiteness. Formula \eqref{A-1-1}
clearly remains true if $Y_+$ is replaced with $Y_-$.)
Consider first $z=1+\xi_N\in(1-\delta,1]$ for $\xi\in(-\delta\alpha_NN^{2/3},0]$.
By \cite[(4.2), (4.6), (4.22)]{DKMVZ2} we have for $S$, the solution
of the Riemann--Hilbert problem [ibid., (4.24)--(4.26)],
(cf.~[ibid., (7.46),(7.47)])
\be\la{M.2.1}
\ba
 S_+(z)=&c_N^{-N\sigma_3}
               \,e^{-\frac{Nl}{2}\sigma_3}\, Y_+(c_Nz+d_N)\\
              &\,\times e^{-N(g_+(z)-\frac{l}{2})\sigma_3}\,
                 \twotwo{1}{0}{-e^{-N(g_+(z)-g_-(z))}}{1},
\ea
\ee
where $\sigma_3=\twotwo{1}{0}{0}{-1}$ and the constant $l\equiv l_N$
is given by [ibid., (5.35)].
Solving for $Y_+$ and substituting in \eqref{A-1-1} we find for
$\xi,\eta\in(-\delta\alpha_NN^{2/3},0]$
\be\la{A-3-1}
\ba
   \frac{c_N}{\alpha_NN^{2/3}}
       &K_N(\xi^{(N)},\eta^{(N)})
   = - \frac{e^{-\frac{N}2(V_N(1+\xi_N)+V_N(1+\eta_N))}}
           {2\pi i(\xi-\eta)}\\
      &\times \onetwo{e^{N(g_+(1+\eta_N)-\frac{l}2)}}
                    {e^{N(g_-(1+\eta_N)-\frac{l}2)}}
      \cdot S_+^T(1+\eta_N)\\
   &\times S_+^{-T}(1+\xi_N)
           \cdot
       \twoone{-e^{N(g_-(1+\xi_N)-\frac{l}2)}}
                    {e^{N(g_+(1+\xi_N)-\frac{l}2)}}.
\ea
\ee
Now note that for $z\in(1-\delta,1]$,
by [ibid., (7.46), (7.47)], $S(z)=R(z)P_N(z)$.
By [ibid., (7.24), (7.9), (7.23), (7.4)],
$$
\ba
 P_{N,+}(z) = \sqrt{\pi}e^{i\pi/6} &\twotwo{1}{-1}{-i}{-i}
            \twotwo{F_N(z)}{0}{0}{1/F_N(z)}\\
       &\times AI_+(f_N(z)) e^{-i\pi\sigma_3/6}
            \twotwo{1}{0}{-1}{1} e^{{N}\Xi_N(z)\sigma_3/2}
\ea
$$
where
$$
       AI(f_N(z))\equiv \twotwo{{\Ai(f_N(z))}}{{\Ai(\omega^2f_N(z))}}
                     {{\Ai^\prime(f_N(z))}}{{\omega^2\Ai^\prime(\omega^2f_N(z))}},
          \qquad \omega=e^{2\pi i/3}.
$$
For $z\in(-1,1)$, in view of [ibid., (5.38)]
$$
          -V_N(z) + g_+(z) + g_-(z) -l = 0
$$
and we find from \eqref{A-3-1}
\be\la{A-4pp-1}
\ba
   \frac{c_N}{\alpha_NN^{2/3}}
       &K_N(\xi^{(N)},\eta^{(N)})
   = -\frac{e^{-\pi i/3}}
           {2\pi i(\xi-\eta)}\,
       \onetwo{1}{0}\cdot AI_+^T(f_N(1+\eta_N))   \\
       &\times     \twotwo{F_N(1+\eta_N)}{0}{0}{1/F_N(1+\eta_N)}
               \twotwo{1}{-i}{-1}{-i}
      R_+^T(1+\eta_N) \\
    &\times R_+^{-T}(1+\xi_N)
          {\twotwo{1}{-i}{-1}{-i}}^{-1}
         \twotwo{1/F_N(1+\xi_N)}{0}{0}{F_N(1+\xi_N)} \\
    &\times
              AI_+^{-T}(f_N(1+\xi_N))    \cdot
       \twoone{0}{1},\qquad \xi,\eta\in(-\delta\alpha_NN^{2/3},0].
\ea
\ee
Similar calculations for $z\in[1,1+\delta)$
lead to the {\em same\ }formula for all other cases $\xi<0,\eta>0$, etc.,
$|\xi|,|\eta|\leq\delta\alpha_NN^{2/3}$.

Now writing
\be\la{resequ}
   R^T(1+\eta_N) = R^T(1+\xi_N)
      +(\eta_N-\xi_N)\int_0^1 (R^T)^\prime\big(1+\xi_N
             +t(\eta_N-\xi_N)\big)\,dt
\ee
and taking into account $\det AI_+(f_N(1+\xi_N))= -1/(2\pi ie^{i\pi/3})$
(use [ibid., (8.38)])
we obtain from \eqref{A-4pp-1} that
\be\la{A-4pp-2}
\ba
   \frac{c_N}{\alpha_NN^{2/3}}
       &K_N(\xi^{(N)},\eta^{(N)})
   =  \frac1{\alpha_NN^{2/3}}Q_{1,1}(\xi,\eta) \\
      &+ \frac1{\xi-\eta}\bigg\{
        \Ai(g_N(\xi)) \Aip (g_N(\eta))\cdot
             \frac{F_N(1+\xi_N)}
               {F_N(1+\eta_N)}- (\xi\leftrightarrow \eta)\bigg\}
\ea
\ee
where $Q_{1,1}$ is as in \eqref{QQ2}.
Now
\be
\la{FNprop2}
\ba
   \frac{F_N(1+\xi_N)}
                {F_N(1+\eta_N )}
      &=\frac{\Fh_N(1+\xi_N)}
                {\Fh_N(1+\eta_N)}\\
  \null  &= 1 + (\xi_N-\eta_N)
       \,\frac{\int_0^1 \Fh_N^\prime(1+\eta_N
                      +  t(\xi_N-\eta_N))\,dt}
               {\Fh_N(1+\eta_N)}
\ea
\ee
and hence using \eqref{eqKA} we rewrite \eqref{A-4pp-2} as
\be\la{A-4pp-3}
\ba
   \frac{c_N}{\alpha_NN^{2/3}}
       K_N(\xi^{(N)},\eta^{(N)})
   =  &\frac{g_N(\xi)-g_N(\eta)}{\xi-\eta}
         \,\KA\big(g_N(\xi),g_N(\eta)\big) \\
     &+ \frac1{\alpha_NN^{2/3}}\big(Q_{1,1}(\xi,\eta) +Q_{1,2}(\xi,\eta)\big)
\ea
\ee
where $Q_{1,2}$ is as in \eqref{Q12}.
Next we write
$
   \frac{g_N(\xi)-g_N(\eta)}{\xi-\eta}
           = \int_0^1 g_N^\prime(\eta+\tau(\xi-\eta))\,d\tau,
$
and use \eqref{gNderiv} and
$$
       \fh_N\big(1+\eta_N+\tau(\xi_N-\eta_N)\big)
       = \fh_N(1) + (\eta_N +\tau(\xi_N-\eta_N))
                  \int_0^1 \fh_N^\prime\big(1+\sigma
          (\eta_N+\tau(\xi_N-\eta_N))\big)\,d\sigma
$$
to conclude that $\frac{g_N(\xi)-g_N(\eta)}{\xi-\eta}=1+E_N(\xi,\eta)$
from \eqref{Q13}. Hence recalling \eqref{eqKA} we obtain from \eqref{A-4pp-3}
\be\la{A-4pp-4}
\ba
   \frac{c_N}{\alpha_NN^{2/3}}
       K_N(\xi^{(N)},\eta^{(N)})
   =  \KA\big(g_N(\xi),g_N(\eta)\big)
                + \frac1{\alpha_NN^{2/3}} \sum_{j=1}^3 Q_{1,j}(\xi,\eta)
\ea
\ee
where $Q_{1,3}$ is as in \eqref{Q13}.
Finally again using \eqref{eqKA} we find
\be\la{KAprop1}
\ba
   \KA(g_N(\xi),g_N(\eta))
      &=   \int_0^\infty \Ai(z+g_N(\xi))\Ai(z+g_N(\eta))\,dz\\
         &=\int_0^\infty \Ai(z+\xi)\Ai(z+\eta)\,dz\\
       &\quad+ \int_0^\infty \Ai(z+\xi)\big[\Ai(z+g_N(\eta))-\Ai(z+\eta)\,dz \\
        &\quad + \int_0^\infty \big[\Ai(z+g_N(\xi))-\Ai(z+\xi)\big]
                       \Ai(z+g_N(\eta))\,dz.
\ea
\ee
The first integral equals $\KA(\xi,\eta)$.
 To evaluate the third integral we recall \eqref{C-1-0}, \eqref{Q14} and note that
\be\la{eq33.p}
\ba
    g_N(\xi)-\xi &= \xi\Big[\fh_N(1+\xi_N) - \fh_N(1)\Big]
\\      &
   = \frac{\xi^2}{\alpha_N N^{2/3}}
    \cdot\int_0^
         1\fh_N^\prime(1+\sigma\xi_N)\,d\sigma
       = \frac{\xi^2}{\alpha_N N^{2/3}} L_N(\xi)
\ea
\ee
which implies
\be\la{Aiprop1}
\ba
   \Ai(z+g_N(\xi))-\Ai(z+\xi) &= \frac{\xi^2}{\alpha_N N^{2/3}}
    \cdot\int_0^
         1\fh_N^\prime(1+\sigma\xi_N)\,d\sigma\\
     &\qquad\qquad\times \int_0^1 \Ai^\prime(z+\xi+\tau(g_N(\xi)-\xi))\,d\tau\\
      &=\frac{\xi^2}{\alpha_N N^{2/3}} L_N(\xi)U_N(\xi,z).
\ea
\ee
The second integral in \eqref{KAprop1} is treated analogously.
We conclude from \eqref{A-4pp-4}, \eqref{KAprop1} that
\eqref{D-11-1} holds
where $Q_{1,4}$ is as in \eqref{Q14}.
The proof of Proposition \ref{AiAi} is complete.
\end{proof}
Now we prove the estimate \eqref{goalCDandderivs}
for $\xi,\eta\in I_N$.
 Note that
by \eqref{gNprop1} it follows that $g_N(\xi), g_N(\eta)$ are bounded
below by some constant $M_0$, and hence
in the region $(\xi,\eta)\in I_N\times I_N$,
 both variables
are bounded below by the constant $L_0$.
Using in addition \eqref{gNprop} we conclude that we can
always use the exponenial bounds on $\Ai$ and its derivatives in \eqref{C-3-1},
and hence
for any $m\in\N$ and $k=0,1,2$, as $N\to\infty$
\be\la{estAiry}
     \Big|\xi^m\big(\frac{d}{d\xi}\big)^k \Ai(g_N(\xi))\Big| \leq C(m) e^{-c(m)\xi},
       \qquad \xi\in I_N.
\ee
Consider $Q_{1,1}(\xi,\eta)$ first.
Recall from \eqref{eq29pp.1} that, in particular,
$\frac{d^j}{dz^j}R(z)=O(N^{-1})$, $j=1,2,3$, uniformly
for $|z-1|\leq \delta$.
It follows then by \eqref{QQ2} using \eqref{eq5.4p},
 \eqref{FNprop}, \eqref{estAiry}
that for $j,k=0,1$
\be
\la{C-3-2}
   \big|\partial_\xi^j\partial_\eta^k
          Q_{1,1}(\xi,\eta)\big| \leq \const \cdot N^{-4/3}e^{-c\xi}e^{-c\eta}
\ee
uniformly for $\xi,\eta\in I_N$.
In the same way we find that for $j,k=0,1$ and $l=2,3,4$
\be
\la{C-3-2prime}
   \big|\partial_\xi^j\partial_\eta^k
          Q_{1,l}(\xi,\eta)\big| \leq \const \cdot N^{-2/3}e^{-c\xi}e^{-c\eta}
\ee
uniformly for $\xi,\eta\in I_N$.
In estimating $Q_{1,4}$, we use the estimate
$$
    |g_N(\xi)-\xi| \leq C\delta|\xi|,\qquad |\xi|\leq\delta\alpha_N N^{2/3},
$$
which follows from \eqref{eq33.p},
together with the uniform boundedness of $L_N(\xi)$
(see \eqref{eqI8.1}(4)): for $\delta$ sufficiently small
this implies that
\be\la{UNbd}
   |U_N(\xi,z)| \leq C e^{-cz} e^{-c\xi},\qquad \xi\in I_N,\qquad z\geq0,
\ee
with similar estimates for the $\xi$- (and $z$-) derivatives.
This proves \eqref{goalCDandderivs} for $(\xi,\eta)\in I_N\times I_N$.
\subsubsection{Estimates for $(\xi,\eta)\in II_N\times II_N$}
\la{ssecEXP}
Recall from [ibid., (4.30), (4.31), (6.16)]
\be\la{eq24p.1}
  S^{(\infty)}(z)\equiv N(z)=
    \frac12\twotwo{a(z)+a(z)^{-1}}{i(a(z)^{-1}-a(z))}
              {i(a(z)-a(z)^{-1})}{a(z)+a(z)^{-1}}
\ee
where $a(z)\equiv\big(\frac{z-1}{z+1}\big)^{1/4}\to1$ as $z\to\infty$.
\begin{proposition}
\la{expexp}
For $j,k=0,1$ and some $C,c>0$
\be
\la{Dp-15-1}
   \bigg|\partial_\xi^j\partial_\eta^k\Big(\frac{c_N}{\alpha_N N^{2/3}}
       K_N(\xin,\etan) \Big)\bigg| \leq C e^{-c N}
             e^{-cN(\xi_N-\delta)}e^{-cN(\eta_N-\delta)}
\ee
uniformly for $\xi,\eta\in II_N$.
\end{proposition}
\begin{proof}
Note first of all that \eqref{A-1-1} still holds.
For  $z=1+\xi_N\in[1+\delta,+\infty)$ we now have
in place of \eqref{M.2.1}
\be\la{M.2.1p}
\ba
 S_+(z)= c_N^{-N\sigma_3}
               \,e^{-\frac{Nl}{2}\sigma_3}\, Y_+(c_Nz+d_N)
        e^{-N(g_+(z)-\frac{l}{2})\sigma_3}
\ea
\ee
where again the constant $l\equiv l_N$
is given by [ibid., (5.35)].
Solving for $Y_+$ and substituting in \eqref{A-1-1} we find for
$\xi,\eta\in II_N$
\be\la{A-3-1p}
\ba
   \frac{c_N}{\alpha_NN^{2/3}}
       K_N(\xi^{(N)},\eta^{(N)})
   = &- e^{-\frac{N}2(V_N(1+\xi_N)-2g_+(1+\xi_N)+l)}
         e^{-\frac{N}2(V_N(1+\eta_N)-2g_+(1+\eta_N)+l)}  \\
        &\times   \frac{\onetwo{1}{0}\cdot S_+^T(1+\eta_N)
                     S_+^{-T}(1+\xi_N)\cdot{\onetwo{0}{1}}^T}
             {2\pi i(\xi-\eta)}.
\ea
\ee
In view of [ibid., (5.38)]
$$
          -V_N(1+\xi_N) + 2g_+(1+\xi_N)  -l = \Xi_{N,+}(1+\xi_N),
    \qquad \xi\in II_N.
$$
Now by [ibid., (2.14), (5.34)] for some $C_1(\delta),C_2(\delta)>0$
and $c>0$ for $N$ large enough
\be\la{expbounds}
\ba
            \Xi_{N,+}(1+\xi_N)   &= -\bigg(\int_1^{1+\delta}
                +\int_{1+\delta}^{1+\xi_N}\bigg)\sqrt{t^2-1}h_N(t)\,dt\\
        & \leq -\int_1^{1+\delta} \sqrt{t^2-1} h_{\textrm{min}}\,dt
                -\int_{1+\delta}^{1+\xi_N} c h_{\textrm{min}}\,dt\\
       & \leq -C_1 -C_2(\xi_N-\delta),\qquad \xi\in(\delta\alpha_NN^{2/3},+\infty).
\ea
\ee
By [ibid., (7.46), (7.47)] for $z\geq1+\delta$,
 $S(z)=R(z)S^{(\infty)}(z)$.
Using \eqref{resequ}, which is still valid for $\xi,\eta\in II_N$, we obtain
\be\la{eq25.p}
\ba
    S_+^T&(1+\eta_N)  S_+^{-T}(1+\xi_N)
  = S_+^{(\infty)T}(1+\eta_N)  S_+^{(\infty)-T}(1+\xi_N) \\
   &+(\eta_N-\xi_N)S_+^{(\infty)T}(1+\eta_N)
        \bigg(\int_0^1 (R_+^T)^\prime\big(1+\xi_N
             +t(\eta_N-\xi_N)\big)\,dt\bigg)
      S_+^{(\infty)-T}(1+\xi_N).
\ea
\ee
Substituting
$$
\ba
    S_+^{(\infty)T}&(1+\eta_N) = S_+^{(\infty)T}(1+\xi_N) \\
    &+(\eta_N-\xi_N)
        \bigg(\int_0^1 (S_+^{(\infty)T})^\prime\big(1+\xi_N
             +t(\eta_N-\xi_N)\big)\,dt
\ea
$$
in the first term in the RHS of \eqref{eq25.p} and noting that
$\onetwo{1}{0}\cdot I\cdot{\onetwo{0}{1}}^T=0$,
we obtain an expression for
  $\onetwo{1}{0}\cdot S_+^{(\infty)T}(1+\eta_N)
 S_+^{(\infty)-T}(1+\xi_N)\cdot{\onetwo{0}{1}}^T$
which is proportional to $(\xi-\eta)$.
The exponential bounds \eqref{Dp-15-1}
then follow from \eqref{expbounds}
and the properties of $S^{(\infty)}$ and $R$
(see \eqref{eq24p.1} and \eqref{resolv}, respectively).
\end{proof}
Now we prove \eqref{goalCDandderivs} for $\xi,\eta\in II_N$
by showing that both of the two terms on the LHS of \eqref{goalCDandderivs}
satisfy the exponential bound.
More precisely, let $\xi\in II_N$. Then either $\xi_N\geq2\delta$
or $\xi_N\in[\delta,2\delta]$. In the former case
\be\la{eqaee2}
   e^{-cN(\xi_N-\delta)} = e^{-cN((\xi_N/2)-\delta)}
         e^{-(cN/(2\alpha_NN^{2/3}))\xi} \leq e^{-\xi},\qquad N\to\infty,
\ee
since $\alpha_N\to(2m)^{2/3}\neq0$ as $N\to\infty$.
In the latter case
$$   e^{-\xi} = e^{-\alpha_NN^{2/3}\xi_N}
         \geq e^{-\alpha_NN^{2/3}2\delta}
       \geq e^{-(c/2)N},\qquad N\to\infty
$$
and hence
\be\la{eqaee3}
    e^{-cN}e^{-cN(\xi_N-\delta)}\leq e^{-(c/2)N}e^{-(c/2)N}
              \leq e^{-(c/2)N}e^{-\xi},\qquad \xi_N\in[\delta,2\delta].
\ee
Combining \eqref{eqaee2} and \eqref{eqaee3}
 we conclude that Proposition \ref{expexp}
implies
\be\la{eqaee1}
 \bigg|\partial_\xi^j\partial_\eta^k\Big(\frac{c_N}{\alpha_N N^{2/3}}
       K_N(\xin,\etan) \Big)\bigg| \leq C e^{-c^\prime N}
             e^{-\xi}e^{-\eta},\qquad \xi,\eta \in II_N.
\ee
Now we consider $\KA(\xi,\eta)$ for $\xi,\eta\in II_N$.
It follows from \cite{Stegun} that
\be\la{eq23.p}
      |\Ai(x)|,|\Ai^\prime(x)|\leq C(L_0) e^{-c(L_0)|x|^{3/2}},\qquad x\geq L_0.
\ee
Using the integral representation \eqref{eqKA}
we estimate for $\xi,\eta\in II_N$
\be\la{eqeaa33}
  |\KA(\xi,\eta)| \leq C\int_0^\infty e^{-c(z+\xi)^{3/2}}e^{-c(z+\eta)^{3/2}}\,dz.
\ee
Let $\xi\in II_N$. Then $\xi\geq1$ for large $N$.
It is elementary to verify that
\be\la{eqaee4}
      (z+\xi)^{3/2} \geq z^{3/2} + \xi^{3/2}, \qquad z\geq 0,\qquad \xi\geq1.
\ee
Next, $\xi^{3/2}\geq (\delta\alpha_N)^{3/2}N\geq \tilde{c}N$, $N\to\infty$
and hence
\be\la{eqaee5}
  \xi^{3/2} - \xi = \xi^{3/2}(1-\xi^{-1/2})
                 \geq c^{\prime\prime} N,\qquad N\to\infty.
\ee
Inserting \eqref{eqaee4}, \eqref{eqaee5} and their analogs for $\eta$
in \eqref{eqeaa33} we find
\be\la{eq24.pp}
  |\KA(\xi,\eta)| \leq C e^{-cN} e^{-c\xi}e^{-c\eta},
                    \qquad\xi,\eta\in II_N.
\ee
A similar argument using \eqref{eq23.p}
also shows that the derivatives of $\KA$ satisfy the same bound.
Combining \eqref{eqaee1} and \eqref{eq24.pp} completes the proof
of \eqref{goalCDandderivs} for $\xi,\eta\in II_N$.
\subsubsection{The ``mixed'' neighborhoods of the end point $1$:
$(\xi,\eta)\in (I_N\times II_N)\cup(II_N\times I_N)$}
\la{ssecMIXED}
Let us consider the case $(\xi,\eta)\in I_N\times II_N$
(the other case is treated analogously).
For $\KA$, we use the bound in \eqref{C-3-1}
for $\xi$,
$$
     |\Ai(z+\xi)|, |\Ai^\prime(z+\xi)|\leq C(L_0)e^{-z}e^{-\xi},
       \qquad z\geq0,\quad\xi\geq L_0,
$$
together with the bound \eqref{eq23.p} for $\eta$.
Inserting these bounds in \eqref{eqKA}
we obtain for $j,k=0,1$
\be\la{eqeaa35}
  |\partial_\xi^j\partial_\eta^k\KA(\xi,\eta)| \leq
         C e^{-cN} e^{-c\xi}e^{-c\eta},
                    \qquad(\xi,\eta)\in I_N\times II_N
\ee
as before. For $K_N(\xin,\etan)$,
there are two cases:
$|\xi_N-\eta_N|\leq\delta/2$ and $|\xi_N-\eta_N|>\delta/2$.
In the first case we can treat
both points as lying in a $I_N\times I_N$ region
corresponding to a larger (fixed) value of $\delta$ (more precisely,
set $\delta\to 3\delta/2$) and hence \eqref{goalCDandderivs}
follows by the arguments in Subsection~\ref{ssecAiAi}.

It remains to consider
the case $(\xi,\eta)\in I_N\times II_N$, $|\xi_N-\eta_N|\geq\delta/2$.
For such $\xi,\eta$, we have
\be\la{eq22p}
     |\xi-\eta|^{-1}\leq N^{-2/3}\alpha_N^{-1}2\delta^{-1}.
\ee
The computations that led to \eqref{A-4pp-1} and \eqref{A-3-1p}
now imply for $\xi\in I_N,\eta\in II_N$
\be
\ba
   \frac{c_N}{\alpha_NN^{2/3}}
       &K_N(\xi^{(N)},\eta^{(N)})
   = - \frac{e^{-\pi i/6}e^{-N\Xi_N(1+\xi_N)/2}}
                  {2\pi i(\xi-\eta)}  \\
        &\times  \onetwo{1}{0}\cdot S^{(\infty)T}(1+\eta_N)
         R_+^T(1+\eta_N)R_+^{-T}(1+\xi_N)\\
 &\times
     {\twotwo{1}{-i}{-1}{-i}}^{-1}
         \twotwo{1/F_N(1+\xi_N)}{0}{0}{F_N(1+\xi_N)} \\
    &\times
              AI_+^{-T}(f_N(1+\xi_N))    \cdot
              \twoone{0}{1},
\ea
\ee
and using the preceding estimates
we find for $j,k=0,1$
$$
      \Big|\partial_\xi^j\partial_\eta^k\frac{c_N}{\alpha_NN^{2/3}}
          K_N(\xi^{(N)},\eta^{(N)}) \Big|
        \leq C N^{-2/3}N^{1/6} e^{-cN}
                     e^{-c\xi} e^{-\eta}
       \leq C  e^{-c^\prime N}
                     e^{-c\xi} e^{-\eta}
$$
uniformly for $(\xi,\eta)\in I_N\times II_N$, $|\xi_N-\eta_N|>\delta/2$.

There is a similar estimate for $(\xi,\eta)\in II_N\times I_N$
which, together with \eqref{eqeaa35}, then proves \eqref{goalCDandderivs}
for $(\xi,\eta)\in (I_N\times II_N)\cup(II_N\times I_N)$.
\subsection{Estimates on integrals of the CD kernel}
For $\xi,\eta\in [L_0,+\infty)$,
making a change
of variables $s=c_N (1+t_N)+d_N$,
and using \eqref{goalCDandderivs} with $j=k=0$,
we readily find
\be\la{D-13-0}
\ba
    \bigg| - &\int_{\xin}^{\infty} K_N(s,\etan)\,ds
    -\Big(-\int_\xi^\infty \KA(t,\eta)\,dt \Big)
  \bigg| \leq C N^{-2/3} e^{-c\xi}e^{-c\eta}\\
    \bigg| - &\int_{\xin}^{\etan} K_N(s,\etan)\,ds
    -\Big(-\int_\xi^\eta \KA(t,\eta)\,dt \Big)
  \bigg| \leq C N^{-2/3} e^{-c\min(\xi,\eta)}e^{-c\eta}
\ea
\ee
uniformly for $\xi,\eta\in [L_0,+\infty)$.
\section{The contribution of the correction term for $\beta=1$ and $4$}
\la{sectfour}
\subsection{Auxiliary facts concerning integrals of the
orthogonal functions $\phi_j$}
It was shown in \cite[(4.14)]{DG} that for a fixed $j\in\N$ the following
holds as $N\to\infty$ (see \eqref{eq5p1})
\be\la{T-14-1}
\ba
    \int_{-\infty}^{+\infty}\phi_{N+j}(y)\,dy
     =c_N^{1/2}\,N^{-1/2}\,(2m)^{-1/2}\,(1+(-1)^{N+j}
               +O(N^{-1/(2m)}))
\ea
\ee
where $2m=\deg V$. Introduce the following column vectors of size $2m-1$
\be\la{eq25.pp}
\ba
   \mathbf{a}&\equiv (1,0,1,0,\cdots,1)^T,\qquad
   \mathbf{b}\equiv (0,1,0,1,\cdots,0)^T\\
   \mathbf{e}&\equiv \mathbf{a}+\mathbf{b}=(1,1,1,\cdots,1)^T.
\ea
\ee
By \eqref{T-1-00} and \eqref{T-14-1} as (even) $N\to\infty$
\be\la{T-14-2}
\ba
   \eps\Phi_1(+\infty)= \frac12\int_{-\infty}^{+\infty}\Phi_1(y)\,dy
     &=c_N^{1/2}\,N^{-1/2}\,(2m)^{-1/2}\,(\mathbf{b}+o(1))\\
   \eps\Phi_2(+\infty)=\frac12\int_{-\infty}^{+\infty}\Phi_2(y)\,dy
     &=c_N^{1/2}\,N^{-1/2}\,(2m)^{-1/2}\,(\mathbf{a}+o(1)).
\ea
\ee
We need also the following result.
Recall the notation \eqref{xinetan}, \eqref{xenn}.
\begin{proposition}
\la{propphiintphi}
For any fixed $j\in\N$ there exist $C,c>0$
such that the following holds as $N\to\infty$,
\be\la{G-4-3}
   \bigg| \phi_{N+j}\big(t^{(N)}\big)
          - \frac{\alpha_N^{1/4}N^{1/6}2^{1/4}}{c_N^{1/2}}\Ai(t)\bigg|
        \leq C c_N^{-1/2}N^{-1/6}e^{-ct},\qquad t\in I_N\cup II_N.
\ee
This estimate implies that for a fixed $j\in\Z$ there exist $C,c>0$ such that
\be\la{T-8-1}
   \bigg| \int_{\xin}^\infty \phi_{N+j}(s)\,ds
          - \frac{c_N^{1/2}}{N^{1/2}}
             \frac{2^{1/4}}{\alpha_N^{3/4}}
              \int_\xi^\infty\Ai(t)\bigg|
        \leq Cc_N^{1/2}N^{-5/6}e^{-c\xi},\qquad \xi\in I_N\cup II_N.
\ee
\end{proposition}
\begin{proof}
Assume first that $j=0$.
It was shown in \cite[Thm.~2.2]{DKMVZ2} 
that
(in our notation)
\be\la{G-2-0}
\begin{aligned}
    \phi_N(t^{(N)})
=  c_N^{-1/2}\bigg[
          & \alpha_N^{1/4}N^{1/6} \hat{F}_N(1+t_N) \Ai(g_N(t))
                            \,\big(1+O(N^{-1})\big)\\
         &- \alpha_N^{-1/4}N^{-1/6} (\hat{F}_N(1+t_N))^{-1} \Ai^\prime(g_N(t))
                            \,\big(1+O(N^{-1})\big)\bigg]
\end{aligned}
\ee
where the error terms are uniform for $t\in I_N$.
Using \eqref{estAiry} we immediately estimate the second term above by
$Cc_N^{-1/2}N^{-1/6}e^{-ct}$ uniformly for $t\in I_N$.
The part of the first term that corresponds to $O(N^{-1})$ is estimated similarly
by $Cc_N^{-1/2}N^{-5/6}e^{-ct}$, $t\in I_N$.
To estimate the leading part of the first term we write
$$
\ba
   \hat{F}_N &(1+t_N) \Ai(g_N(t)) =  \hat{F}_N(1) \Ai(t)\\
                &+ \hat{F}_N(1) \big[ \Ai(g_N(t))-\Ai(t)\big]
                 + \Ai(g_N(t))\big[ \hat{F}_N(1+t_N) - \hat{F}_N(1) \big].
\ea
$$
By \eqref{eqI8.1}(5) and \eqref{C-1-0},  $\hat{F}_N(1)=2^{1/4}$.
By formula \eqref{Aiprop1} and \eqref{UNbd}
$$
    \big|\Ai(g_N(t)) - \Ai(t)\big|
            \leq CN^{-2/3}t^2e^{-ct}\leq C^\prime N^{-2/3}e^{-c^\prime t},
                  \qquad t\in I_N.
$$
Also using \eqref{estAiry} and \eqref{FNprop} we obtain
$$
       \big|\Ai(g_N(t))\big|\cdot\big|\hat{F}_N(1+t_N)-\hat{F}_N(1)\big|
       \leq CN^{-2/3}|t|e^{-ct}\leq C^\prime N^{-2/3}e^{-c^\prime t},
                  \qquad t\in I_N.
$$
Combining the above estimates we find that
$$
   \big| \hat{F}_N (1+t_N) \Ai(g_N(t)) -  2^{1/4} \Ai(t)\Big|
             \leq C N^{-2/3} e^{-c t},\qquad t\in I_N,
$$
which completes the proof of \eqref{G-4-3} for $j=0$ and $t\in I_N$.
We now consider \eqref{G-4-3} for $j=0$
and $t\in II_N=[\delta\alpha_NN^{2/3},\infty)$.
For such $t$, by \eqref{C-3-1},
$$
 |\Ai(t)| \leq Ce^{-t} \leq C e^{-cN^{2/3}}e^{-t/2}
$$
and hence
$
   \big|\frac{\alpha_N^{1/4}N^{1/6}2^{1/4}}{c_N^{1/2}}\Ai(t)\big|
        \leq C c_N^{-1/2}N^{-1/6}e^{-ct} 
$.
Also from \cite[(4.8)]{DG},
we find
$
    \big|\phi_{N}(t^{(N)})\big|
         \leq C c_N^{-1/2} e^{-cN} e^{-ct} 
$.
These two estimates for $t\in II_N$, together with the previous estimate
for $t\in I_N$, yield \eqref{G-4-3} in the case $j=0$
for all $t\in[L_0,\infty)$.

Now fix any $j\in\Z$
and write
\be\la{G-5pp-1}
    \phi_{N+j}\bigg(c_N\Big(1+\frac{t}{\alpha_NN^{2/3}}\Big)  +d_N\bigg)
         = \phi_{N+j} \bigg( c_{N+j}\Big(1
             +\frac{t_{N,j}}{\alpha_{N+j}(N+j)^{2/3}}\Big)
                + d_{N+j} \bigg)
\ee
where
\be\la{eq26.p}
\ba
  t_{N,j} &= t \cdot \frac{c_N}{c_{N+j}}
                                \frac{\alpha_{N+j}}{\alpha_{N}}
                                 \frac{(N+j)^{2/3}}{N^{2/3}}
               + \bigg(\frac{c_N}{c_{N+j}}-1\bigg)\cdot \alpha_{N+j}(N+j)^{2/3}
               \\ &\qquad\qquad+ \frac{d_{N+j}}{c_{N+j}}
                         \bigg(\frac{d_N}{d_{N+j}}-1\bigg)\cdot \alpha_{N+j}(N+j)^{2/3}\\
             &= (1+O(N^{-1}))\,t + O(N^{-1/3})
\ea
\ee
by \eqref{cN}, \eqref{dN}, \eqref{eqI8.1}(2).
In particular, as $N\to\infty$,
  $t_{N,j}\geq(1-\frac12\sgn L_0)L_0$.
Now the RHS of \eqref{G-5pp-1} can be written as
$\phi_{N^\prime}((t_{N,j})^{(N^\prime)})$ where $N^\prime=N+j$.
Applying the estimate \eqref{G-4-3} just derived for $j=0$,
with $L_0$ replaced by $(1-\frac12\sgn L_0)L_0$,
   we obtain
\be\la{eq27p.1}
   \bigg| \phi_{N^\prime}\big((t_{N,j})^{(N^\prime)}\big)
          - \frac{\alpha_{N+j}^{1/4}(N+j)^{1/6}2^{1/4}}{c_{N+j}^{1/2}}
               \Ai(t_{N,j})\bigg|
        \leq C c_{N+j}^{-1/2}(N+j)^{-1/6}e^{-ct_{N,j}}
\ee
for all $t\geq L_0$.
Using \eqref{eq26.p}, and also \eqref{cN}, \eqref{eqI8.1}(2),
together with the elementary estimate
$$
       \big|\Ai(t_{N,j}) -\Ai(t)\big|
              \leq C^\prime N^{-1/3} e^{-c^\prime t}
$$
(use \eqref{C-3-1}), we obtain \eqref{G-4-3} from \eqref{eq27p.1}
for any fixed $j\in\Z$.

Finally \eqref{T-8-1} follows readily by integrating \eqref{G-4-3}.
\end{proof}
Recall the notation \eqref{eq25.pp}.
Proposition \ref{propphiintphi} implies that for $j=1,2$ one has
uniformly for $t,\xi,\eta\geq L_0$
\be\la{PHIsumm}
\ba
      \bigg| &\Phi_j\big(t^{(N)}\big)
          - \frac{\alpha_N^{1/4}N^{1/6}2^{1/4}}{c_N^{1/2}}\Ai(t)
           \,\mathbf{e}\bigg|
        \leq Cc_N^{-1/2}N^{-1/6}e^{-ct}\\
     \bigg| &\int_{\xin}^\infty \Phi_j(s)\,ds - \bigg(\frac{c_N^{1/2}}{N^{1/2}}
             \frac{2^{1/4}}{\alpha_N^{3/4}}
              \int_\xi^\infty\Ai(t)\,dt\bigg)\mathbf{e}
                           \bigg| \leq Cc_N^{1/2}N^{-5/6}e^{-c\xi}\\
     \bigg| &\int_{\xin}^{\etan} \Phi_j(s)\,ds - \bigg(\frac{c_N^{1/2}}{N^{1/2}}
             \frac{2^{1/4}}{\alpha_N^{3/4}}
              \int_\xi^\eta\Ai(t)\,dt\bigg)\mathbf{e}\bigg|
          \leq Cc_N^{1/2}N^{-5/6}e^{-c\min(\xi,\eta)}.
\ea
\ee
\subsection{The case $\beta=4$}
\subsubsection{The contribution of the correction term to
the $12$ entry of $K_{N,4}$}
\la{ssec12beta4}
Since $(SD)(x,y)=-\partial_y S(x,y)$,
the correction term in \eqref{W4} has
the form
\be\la{T-5-15}
   -\Phi_2(x)^T\cdot D_{21}\cdot\Phi_1(y)
          -\Phi_2(x)^T\cdot D_{21}C_{11}^{-1}B_{11}D_{12}\cdot\Phi_2(y).
\ee
Set $x=\xin$, $y=\etan$.
Recall $n=2m-1$, $2m=\deg V$.
Note that by \cite[(2.13)]{DG}
\be\la{T-5-4}
 D_{21} =
     \frac{m\kappa_{2m}}{2^{2m-1}}\,c_N^{2m-1}\,
\left[\begin{pmatrix}
&\binom{n}{0} &0&\binom{n}{1}&\cdots&\binom{n}{(n-1)/2} \\
&  0 &1&0&\cdots&0\\
&\cdots\\
&0 &0&0&\cdots&1
\end{pmatrix} +o(1)\right]
\ee
and $D_{21}=O(c_N^{2m-1})=O(N^{1-1/(2m)})$
as $N\to\infty$ (see \eqref{cN}). Also
since $C_{11}=I-B_{12}D_{21}$, we see from [ibid., (2.19)]
that $C_{11}^{-1}B_{11}$ is skew symmetric,
being the lower right corner of the skew symmetric matrix $D_N^{-1}$.
(Note that $B_{11}$ is the lower right $n\times n$ corner of $\eps_N$.)
Hence $D_{21}C_{11}^{-1}B_{11}D_{12}$ in \eqref{T-5-15}
is also skew, and using [ibid., (2.13)] and the fact that
$C_{11}^{-1}$ is bounded as $N\to\infty$ [ibid., Thm.~2.4, 2.6],
we see that
   $D_{21}C_{11}^{-1}B_{11}D_{12}=O(N^{1-1/(2m)})$ as $N\to\infty$.
Recall that the $12$ entry in $K_{N,4}(\xin,\etan)$ has an overall scaling factor
$\big(\frac{c_N}{\alpha_NN^{2/3}}\big)^2$.
Substituting the leading term in the representation of $\Phi_j$ in
\eqref{PHIsumm} into the first term in \eqref{T-5-15},
and using \eqref{T-5-4}, we obtain
\be\la{T-6-1}
\ba
  - \frac{c_N^2}{\alpha_N^2N^{4/3}}\,
      \frac{m\kappa_{2m}}{2^{2m-1}}\,c_N^{2m-1}\,
       \frac{\alpha_N^{1/2}N^{1/3}2^{1/2}}{c_N} (\Sigma_n +o(1))
               \Ai(\xi)\Ai(\eta)
\ea
\ee
where $o(1)$ is independent of $\xi,\eta$
and $\Sigma_n$ denotes the sum of all elements of the first (binomial)
matrix on the RHS in \eqref{T-5-4}.
Using the formula preceding [ibid., (6.7)] one finds
$$
   \Sigma_n=\frac12\frac{m(2m)!}{(m!)^2}.
$$
Recall that by \eqref{eqI8.1}(2), [ibid., (2.14)] and \eqref{C-3-1},
\be\la{Airyagain}
\ba
     \null &\alpha_N=2m^{2/3}+o(1),\qquad
      \frac{c_N^{2m}}{N}
\frac{m\kappa_{2m}}{2^{2m-1}}
   = \frac{2(m!)^2}{(2m)!} + o(1),\qquad N\to\infty\\
   \null&|\Ai(\xi)|\leq Ce^{-\xi},\qquad \xi\geq L_0.
\ea
\ee
Inserting these estimates, \eqref{T-6-1} becomes
$$
  - \frac12 \Ai(\xi)\Ai(\eta)  + o(1)e^{-\xi}e^{-\eta}
$$
uniformly for $\xi,\eta\geq L_0$
and $o(1)$ is independent of $\xi,\eta$.
The error that was made by substituting only the leading term in
\eqref{PHIsumm} in the first term in \eqref{T-6-1}, is estimated as follows:
\be\la{T-6-1err}
\ba
    O(c_N^2 N^{-4/3})\,O(c_N^{2m-1})\,
       \Big\{ &O(N^{1/6}{c_N^{-1/2}}) c_N^{-1/2}N^{-1/6}\big(
               |\Ai(\xi)| e^{-c\eta} + |\Ai(\eta)| e^{-c\xi} \big)\\
       &\qquad+ O(N^{-1/3}{c_N^{-1}})   e^{-c\xi} e^{-c\eta}  \Big\}\\
   &\leq O(N^{-1/3}) e^{-c\xi} e^{-c\eta}
\ea
\ee
uniformly for $\xi,\eta\geq L_0$, and independent of the degree $2m$ of $V$.

Next we substitute the leading terms in the representation of $\Phi_2$ in
\eqref{PHIsumm} in the second term in \eqref{T-5-15}.
By the skew symmetry of  $D_{21}C_{11}^{-1}B_{11}D_{12}$ noted above,
the result is precisely zero.
The error that is made by such a substitution
is estimated in exactly the same way as in \eqref{T-6-1err} and is
also of order
\be\la{T-6-1p}
\ba
     O(N^{-1/3})e^{-c\xi}e^{-c\eta}
\ea
\ee
uniformly for $\xi,\eta\geq L_0$.
We conclude that the contribution of the correction term
to the $12$ entry is given by
\be\la{T-7-0}
  - \frac12 \Ai(\xi)\Ai(\eta)  + o(1)e^{-c\xi}e^{-c\eta}
\ee
uniformly for $\xi,\eta\geq L_0$.
\subsubsection{The contribution of the correction term to
the $11$ and $22$ entries of $K_{N,4}$}
\la{ssec11beta4}
We consider the $11$ entry of $K_{N,4}$ (the $22$ entry
is analyzed in the same way).
The correstion term in \eqref{T-1-1} has the form
\be\la{eq4.10p}
\ba
   \Phi_2(x)^T\cdot &D_{21}\cdot
          \bigg(-\int_{y}^\infty\Phi_1(t)\,dt\bigg)\\
          \null&+\Phi_2(x)^T\cdot D_{21}C_{11}^{-1}B_{11}D_{12}\cdot
           \bigg(-\int_{y}^\infty\Phi_2(t)\,dt\bigg).
\ea
\ee
We set $x=\xin$, $y=\etan$ in \eqref{eq4.10p}.
The $11$ (and $22$) entry in $K_{N,4}(\xin,\etan)$
 has an overall scaling factor
$\frac{c_N}{\alpha_NN^{2/3}}$.
Hence, substituting the leading terms in the representation of $\Phi_2$,
$\int\Phi_1$ in
\eqref{PHIsumm} into the first term in \eqref{eq4.10p}
and using \eqref{T-5-4}, we obtain
\be\la{T-8-2}
\ba
   \frac{c_N}{\alpha_NN^{2/3}}\,
      \frac{m\kappa_{2m}}{2^{2m-1}}\,c_N^{2m-1}\,
       \frac{\alpha_N^{1/4}N^{1/6}2^{1/4}}{c_N^{1/2}}
      \frac{c_N^{1/2}2^{1/4}}{\alpha_N^{3/4}N^{1/2}}
        (\Sigma_n +o(1))
               \Ai(\xi)\bigg(-\int_\eta^\infty\Ai(t)\,dt\bigg)
\ea
\ee
where $o(1)$ is independent of $\xi,\eta$.
Computing the factor and using \eqref{Airyagain}
as above we see that \eqref{T-8-2} becomes
$$
  - \frac12 \Ai(\xi)\int_\eta^\infty\Ai(t)\,dt  + o(1)e^{-\xi}e^{-\eta}
$$
uniformly for $\xi,\eta\geq L_0$
and $o(1)$ is independent of $\xi,\eta$.
The error that was made by substituting only the leading terms
for $\Phi_2$, $\int\Phi_1$ in
\eqref{PHIsumm} into the first
term in \eqref{eq4.10p}, is estimated as follows:
\be\la{T-8-2err}
\ba
    O(c_N N^{-2/3})\,O(c_N^{2m-1})\,
       \bigg( &\frac{N^{1/6}}{c_N^{1/2}}\frac{c_N^{1/2}}{N^{5/6}}
                +\frac{1}{N^{1/6}c_N^{1/2}}\frac{c_N^{1/2}}{N^{1/2}}
               +\frac{c_N^{1/2}}{N^{5/6}}\frac{1}{N^{1/6}c_N^{1/2}}
                   \bigg)e^{-c\xi} e^{-c\eta}\\
       &= O(N^{1/3})
       \big(N^{-2/3}+N^{-2/3}+N^{-1}\big)  e^{-c\xi} e^{-c\eta} \\
   &= O(N^{-1/3}) e^{-c\xi} e^{-c\eta}
\ea
\ee
uniformly for $\xi,\eta\geq L_0$,
and again independent of the degree $2m$ of $V$.

Next we substitute the leading terms in the representation of $\Phi_2$,
$\int\Phi_2$ into
\eqref{PHIsumm} in the second term in \eqref{eq4.10p}.
Again by skew symmetry,
the result is precisely zero.
The error that is made by such a substitution
is estimated in exactly the same way as in \eqref{T-8-2err}
 and also has order
\be\la{T-8-2p}
\ba
     O(N^{-1/3})e^{-c\xi}e^{-c\eta}
\ea
\ee
uniformly for $\xi,\eta\geq L_0$.
We conclude that the
contribution of the correction term to
the $11$ entry is given by
\be\la{T-8-2l}
     - \frac12 \Ai(\xi)\int_\eta^\infty\Ai(t)\,dt  + o(1)e^{-c\xi}e^{-c\eta}
\ee
uniformly for $\xi,\eta\geq L_0$
(for the $22$ entry $\xi$ and $\eta$ should be interchanged).
\subsubsection{The contribution of the correction term to
the $21$ entry of $K_{N,4}$}
By \eqref{T-1-1}, \eqref{1.47beta4} the correction term in the $21$
entry of $K_{N,4}$ is given by
\bq
\label{T-10-1}
\ba
   \int_x^\infty &\Phi_2^T(s)\,ds \cdot D_{21}\cdot
          \int_{y}^\infty\Phi_1(t)\,dt\\
          &\qquad+\int_x^\infty \Phi_2^T(s)\,ds
          \cdot D_{21}C_{11}^{-1}B_{11}D_{12}\cdot
           \int_{y}^\infty\Phi_2(t)\,dt.
\ea
\eq
Again we replace $x=\xin$, $y=\etan$.
Recall that the $21$ entry in $K_{N,4}(\xin,\etan)$ has no overall scaling factor.
Substituting the leading terms in the representation of $\int\Phi_j$,
$j=1,2$, in \eqref{PHIsumm}
into the first term in \eqref{T-10-1}
in the same way as before, we obtain
$$
\ba
      \frac{m\kappa_{2m}}{2^{2m-1}}\,c_N^{2m-1}\,
      &\frac{c_N 2^{1/2}}{\alpha_N^{3/2}N}
        (\Sigma_n +o(1))
               \int_\xi^\infty \Ai(s)\,ds \int_\eta^\infty\Ai(t)\,dt\\
     &=\frac12 \int_\xi^\infty \Ai(s)\,ds
            \int_\eta^\infty\Ai(t)\,dt  + o(1)e^{-\xi}e^{-\eta}
\ea
$$
uniformly for $\xi,\eta\geq L_0$
and $o(1)$ is independent of $\xi,\eta$.
The error just made is estimated as follows:
\be\la{T-11-1err}
\ba
    O(c_N^{2m-1})\,
       \bigg( &2\frac{c_N^{1/2}}{N^{1/2}}\frac{c_N^{1/2}}{N^{5/6}}
               +\frac{c_N}{N^{5/3}}
                   \bigg)e^{-c\xi} e^{-c\eta}\\
       &= O(N^{1/3})
       \big(N^{-4/3}+N^{-5/3}\big)  e^{-c\xi} e^{-c\eta} \\
   &= O(N^{-1/3}) e^{-c\xi} e^{-c\eta}
\ea
\ee
uniformly for $\xi,\eta\geq L_0$,
here all order factors are independent of $\xi,\eta$.
(Here we have used $|\int_\xi^\infty\Ai(t)\,dt|\leq Ce^{-\xi}$
uniformly for $\xi\geq L_0$.)

Finally, we substitute the leading terms in the representation of
$\int\Phi_j$, $j=1,2$, in
\eqref{PHIsumm} into the second term in \eqref{T-10-1}.
By the skew symmetry
the result is again precislely zero.
The error that is made by such a substitution
is estimated in exactly the same way as in \eqref{T-11-1err}
and is also of order
\be\la{T-11-1p}
\ba
     O(N^{-1/3})e^{-c\xi}e^{-c\eta}
\ea
\ee
uniformly for $\xi,\eta\geq L_0$. We conclude that the
contribution of the correction term to
the $21$ entry is given by
\be\la{T-11-1}
   \frac12 \int_\xi^\infty \Ai(s)\,ds
            \int_\eta^\infty\Ai(t)\,dt  + o(1)e^{-c\xi}e^{-c\eta}
\ee
uniformly for $\xi,\eta\geq L_0$.
\subsection{The case $\beta=1$}
As we will see, this case is more involved than the case $\beta=4$.
Consider the
 $2n\times2n$ ($n=2m-1$, $2m=\deg V$)
matrix $(AC(I_{2n}-BAC)^{-1})^T$
in the $\beta=1$ correction term in \eqref{W1}
as a two by two block matrix with blocks
of size $n\times n$. Denote the upper left and the upper right
blocks by $G_{11}$ and $G_{12}$, respectively.
With this notation the correction term has the form
\be\la{T-12p-1}
\ba
     - \Phi_1(x)^T \cdot G_{11} \cdot\eps\Phi_1(y)
       - \Phi_1(x)^T \cdot G_{12} \cdot\eps\Phi_2(y).
\ea
\ee
As in \cite{DG} let $R\equiv{}R_n$
 denote the $n\times n$
matrix with all entries zero apart from
ones on the anti-diagonal (thus $R_{i,j}=1$
if $j=n-i+1$, $1\leq i\leq n$, and $R_{i,j}=0$ otherwise).
Note that $R^2=I_n$.
Define
\be\la{G11tilde}
      \tilde{G}_{11}\equiv - RD_{21}C_{11}^{-1}B_{11}D_{12} R.
\ee
Note from Subsection \ref{ssec12beta4}
 that $D_{21}C_{11}^{-1}B_{11}D_{12}$
is skew and of order $O(N^{1-1/(2m)})$ as $N\to\infty$.
 Hence $\tilde{G}_{11}$ is also skew and has the same order as $N\to\infty$.
We need the following result.
\begin{proposition}
\la{propbeta1}
As (even) $N\to\infty$ we
have $G_{11},G_{12}=O(N^{1-1/(2m)})$, more precisely
\be\la{T-16-1}
     G_{11} =
           \tilde{G}_{11} + o(N^{1-1/(2m)}),\qquad N\to\infty,
\ee
and also
\be\la{T-16-000}
      G_{12} =  D_{12} + o(N^{1-1/(2m)}),\qquad N\ra\infty.
\ee
\end{proposition}
\begin{proof}
It was shown in \cite[Theorem 2.3]{DG} that, as $N\to\infty$,
\bq
\la{BAC}
\ba
    (BA)_{22}&=-R(BA)_{11}R+o(1)\\
      BAC&=\twotwo{0}{0}{(BA)_{21}+o(1)}{(BA)_{22}+o(1)}.
\ea
\eq
Denote
\be\la{eqT}
       T\equiv I_n - (BAC)_{22} = I_n - (BA)_{22}+o(1)
         = I_n +R (BA)_{11}R +o(1)= R C_{11}R +o(1).
\ee
It was shown in \cite[Theorem 2.6]{DG} that,
as $N\ra\infty$, $T$ approaches
a constant nondegenerate matrix.
Thus
$$
     (I_{2n} - BAC)^{-1}
            = \twotwo{I_n}{0}
                             {T^{-1}((BA)_{21}+o(1))}{T^{-1}}
$$
and simple algebra using \eqref{B}, \eqref{A}
 now shows that in the product
$AC(I_{2n} - BAC)^{-1}
           = \twotwo{G_{11}^T}{*}{G_{12}^T}{*}$
we have by \eqref{BAC}
\be\la{V-2-3}
\ba
  G_{11}^T &= A_{12}\big[
                  (BA)_{21} + (BA)_{22}T^{-1}((BA)_{21}+o(1))\big]\\
    G_{12}^T &=A_{21}\big[I_n + (BA)_{11} + (BA)_{12}T^{-1}
                                    ((BA)_{21}+o(1))\big].
\ea
\ee
Using \eqref{eqT}, this implies
\be\la{V-3-1}
\ba
   N^{-1+1/(2m)} G_{11}^T &= N^{-1+1/(2m)} A_{12}\big[
                  I_n + (BA)_{22}T^{-1}\big] (BA)_{21}+o(1)\\
          &= N^{-1+1/(2m)} A_{12}\big[
                  T + (BA)_{22}\big]T^{-1} (BA)_{21}+o(1)\\
     &= N^{-1+1/(2m)} A_{12} T^{-1} (BA)_{21}+o(1).
\ea
\ee
Now from
\be\la{BAsec7}
       BA = \twotwo{B_{12}A_{21}}{B_{11}A_{12}}
                              {B_{22}A_{21}}{B_{21}A_{12}}
\ee
and \eqref{V-3-1}, \eqref{eqT} we obtain
$$
\ba
   N^{-1+1/(2m)} G_{11}^T
       &= N^{-1+1/(2m)} A_{12}
                (RR + R(BA)_{11}R)^{-1} B_{22}A_{21}+o(1)\\
      &= N^{-1+1/(2m)} A_{12}R (I_n + (BA)_{11})^{-1}R B_{22}A_{21}+o(1)\\
     &= N^{-1+1/(2m)} A_{12}R C_{11}^{-1}R B_{22}RRA_{21}+o(1).
\ea
$$
Using the asymptotic relations
\be\la{eq31.p1}
\ba
    N^{-1+1/(2m)} RA_{12}R &= N^{-1+1/(2m)} A_{21} + o(1)\\
    N^{1-1/(2m)} RB_{22}R &= -N^{1-1/(2m)} B_{11} + o(1)
\ea
\ee
from \cite[Subsec.~5.2]{DG} we see that
$$
\ba
   N^{-1+1/(2m)} RG_{11}^TR
      &= N^{-1+1/(2m)} (RA_{12}R) C_{11}^{-1}(R B_{22}R)
                   (RA_{21}R)+o(1)\\
      &= -N^{-1+1/(2m)} A_{21} C_{11}^{-1}B_{11} A_{12}+o(1)\\
      &= N^{-1+1/(2m)} D_{21} C_{11}^{-1}B_{11} D_{12}+o(1).
\ea
$$
As noted above, the matrix $D_{21} C_{11}^{-1}B_{11} D_{12}$
 is skew symmetric and hence
$$
   N^{-1+1/(2m)} G_{11}
   = -N^{-1+1/(2m)}
                  RD_{21} C_{11}^{-1}B_{11} D_{12}R+o(1)
$$
which proves \eqref{T-16-1}.

Now let us prove \eqref{T-16-000}.
From  \eqref{V-2-3} we derive
$$
\ba
    N^{-1+1/(2m)} G_{12}^T
       &= N^{-1+1/(2m)}A_{21}\big[I_n + (BA)_{11} + (BA)_{12}T^{-1}
                                    (BA)_{21}\big]+o(1)
\ea
$$
and hence, because $A_{21}^T=A_{12}=D_{12}$,
 we note that we just have to prove
$$
   N^{-1+1/(2m)}A_{21}\big[ (BA)_{11} + (BA)_{12}T^{-1}
                                    (BA)_{21}\big] = o(1).
$$
Since $N^{-1+1/(2m)}A_{21}=O(1)$,
it is sufficient to prove
$$
    (BA)_{11} + (BA)_{12}T^{-1}
                                    (BA)_{21} = o(1).
$$
By \eqref{BAsec7} the LHS is
$
   B_{12}A_{21} + B_{11}A_{12}T^{-1}B_{22}A_{21}
$
and so we see that it is sufficient to show that
$$
     B_{12} + B_{11}A_{12}T^{-1}B_{22} = o(N^{-1+1/(2m)}).
$$
Using \eqref{eqT} this reduces to showing that
$$
     B_{12} + B_{11}A_{12}RC_{11}^{-1}RB_{22} = o(N^{-1+1/(2m)})
$$
or
$$
        RB_{12}R + (RB_{11}R)(RA_{12}R)C_{11}^{-1}
                  (RB_{22}R) = o(N^{-1+1/(2m)}).
$$
Using $N^{-1+1/(2m)} RB_{12}R = -N^{-1+1/(2m)} B_{21} + o(1)$
which follows as in \eqref{eq31.p1},
we are reduced to proving finally
\be\la{V-4-1}
          -B_{21} + B_{22}A_{21}C_{11}^{-1}B_{11} = o(N^{-1+1/(2m)}).
\ee
But
$$
       -  B_{21} + B_{22}A_{21}C_{11}^{-1} B_{11} = 0
$$
by (taking the transposes of) \cite[(5.12)]{DG}.
The proof of Proposition \ref{propbeta1} is complete.
\end{proof}
\begin{remark}
The second relation in \eqref{BAC} was sharpened recently by
Kriecherbauer and Vanlessen \cite{KV} who showed that
the $o(1)$ terms are in fact identically zero.
One might hope that this improved result could be used to
strengthen the estimates in \eqref{T-16-1}, \eqref{T-16-000}.
This is indeed the case for \eqref{T-16-000}:
one can show that $G_{12}=D_{12}$ identically.
However we have not been able to use \cite{KV}
to improve the estimate in \eqref{T-16-1}.
\end{remark}
\subsubsection{The contribution of the correction term to
the $12$ entry of $K_{N,1}$}
\la{ssec12beta1}
In view of \eqref{T-12p-1}, since $(SD)(x,y)=-\partial_y S(x,y)$,
the correction term has the form
\be\la{T-12-2p}
    \Phi_1(x)^T \cdot G_{11} \cdot\Phi_1(y)
       + \Phi_1(x)^T \cdot G_{12} \cdot\Phi_2(y).
\ee
Again set $x=\xin$, $y=\etan$.
Using Proposition \ref{propbeta1}
 and proceeding in the same way
as in Subsection \ref{ssec12beta4} we find that as $N\to\infty$,
the term \eqref{T-12-2p},
 multiplied as before by $(\frac{c_N}{\alpha_NN^{2/3}})^2$,
becomes
\be\la{T-12-2}
    -\frac12\Ai(\xi)\Ai(\eta) + o(1)e^{-c\xi}e^{-c\eta}
\ee
uniformly for $\xi,\eta\geq L_0$.
Note that the sum of all elements of (the binomial matrix in the limiting form of)
$D_{12}$ is, up to a sign,
the same as for $D_{21}$.

\noindent
\textbf{Remark:}
Note also that the only new element in the above proof as compared
 with the case $\beta=4$
in Subsection \ref{ssec12beta4},
is that the matrix $G_{11}$
is only asymptotically (and not identically) skew symmetric.
This leads to the estimate
$o(1)e^{-c\xi}e^{-c\eta}$ in place of \eqref{T-6-1p}.
\subsubsection{The contribution of the correction term to
the $11$ and $22$ entries of $K_{N,1}$}
\la{ssec4.3.2}
We consider the $11$ entry of $K_{N,1}$ (the $22$ entry
is considered in the same way).
Using \eqref{T-1-00}, \eqref{T-12p-1} we rewrite the correction term as
\be\la{T-13-1}
\ba
   -\Phi_1(x)^T\cdot &G_{11}\cdot \bigg(-\int_y^\infty\Phi_1(t)\,dt\bigg)
          -\Phi_1(x)^T\cdot G_{12} \cdot \bigg(-\int_y^\infty\Phi_2(t)\,dt\bigg)\\
   &-\Phi_1(x)^T\cdot G_{11}\cdot\eps\Phi_1(+\infty)
          -\Phi_1(x)^T\cdot G_{12} \cdot\eps\Phi_2(+\infty).
\ea
\ee
Again set $x=\xin$, $y=\etan$.
The first two terms can be treated in the same way
as in Subsections \ref{ssec11beta4} and \ref{ssec12beta1}.
More precisely we find that the first two terms in \eqref{T-13-1},
multiplied by $\frac{c_N}{\alpha_NN^{2/3}}$,
 become, as $N\to\infty$
\be\la{T-13-2}
    -\frac12\Ai(\xi)\int_\eta^\infty\Ai(t)\,dt + o(1)e^{-c\xi}e^{-c\eta}
\ee
uniformly for $\xi,\eta\geq L_0$.
Now consider the (scaled) sum of the last two terms in \eqref{T-13-1}
\be\la{T-20-1}
     -\frac{c_N}{\alpha_NN^{2/3}}\Big(
            \Phi_1(\xin)^T\cdot G_{11}\cdot\eps\Phi_1(+\infty)
          +\Phi_1(\xin)^T\cdot G_{12} \cdot\eps\Phi_2(+\infty)\Big).
\ee
By \eqref{T-14-2}, \eqref{PHIsumm},
Proposition \ref{propbeta1}, this
becomes as $N\to\infty$
\be\la{T-14-22}
\ba
     -\frac{c_N}{\alpha_NN^{2/3}}
       &\frac{\alpha_N^{1/4}N^{1/6}2^{1/4}}{c_N^{1/2}}
       \frac{c_N^{1/2}}{(2m)^{1/2}N^{1/2}}\\
    &\times\Big\{
            \mathbf{e}^T\cdot G_{11}\cdot (\mathbf{b}+o(1))
          +\mathbf{e}^T\cdot G_{12} \cdot  (\mathbf{a}+o(1))\Big\}
                    \Ai(\xi) + o(1)e^{-c\xi}.
\ea
\ee
Setting $\mathbf{a}=\mathbf{e}-\mathbf{b}$,
we find that \eqref{T-14-22} reduces to
\be\la{T-14-23}
\ba
    \frac12 \Ai(\xi) &+ o(1)e^{-c\xi}
           + \Big(\mathbf{e}^T\cdot  G_{11}\cdot \mathbf{b}
                      -   \mathbf{e}^T\cdot  G_{12}\cdot \mathbf{b}
                   \Big)\cdot\Ai(\xi)\cdot O(N^{-1+1/(2m)}).
\ea
\ee
So if we could prove
\be\la{T-15-1}
   \mathbf{e}^T\cdot G_{11}\cdot \mathbf{b}
         -   \mathbf{e}^T\cdot G_{12}\cdot \mathbf{b}
        = o(N^{1-1/(2m)}),\qquad N\to\infty,
\ee
then we would find that \eqref{T-14-23} equals
\be\la{T-15-1l}
    \frac12 \Ai(\xi) + o(1)e^{-c\xi}
\ee
uniformly for $\xi\geq L_0$.

We prove \eqref{T-15-1}.
We will, perhaps surprisingly,
use a property of the $\beta=4$ correlation kernel $S_{N/2,4}$:
it is not clear how to prove \eqref{T-15-1}
directly using the asymptotic properties of $(D\phi_{N+j},\phi_{N+k})$
and $(\eps\phi_{N+j},\phi_{N+k})$ given in \cite{DG}.
More precisely, \eqref{T-15-1} follows from \eqref{T-16-1}, \eqref{T-16-000}
and the relation
\be\la{T-16-1p}
   \mathbf{b} + C_{11}^{-1}B_{11}D_{12}\cdot \mathbf{a} =
               o(1),\qquad N\to\infty,
\ee
which is proved by dividing \eqref{eq1.42.p}
by $(\frac{c_N}{2mN})^{1/2}$
and using \eqref{T-14-2} as $N\to\infty$.
Multiplying \eqref{T-16-1p}
from the left by $\mathbf{e}^T\cdot RD_{21}$ and
noting $\mathbf{b}=R\cdot \mathbf{b}$, $\mathbf{a}=R\cdot \mathbf{a}$,
 we find
$$
   \mathbf{e}^T\cdot RD_{21}R\cdot \mathbf{b} +
         \mathbf{e}^T\cdot
                  RD_{21}C_{11}^{-1}B_{11}D_{12}R\cdot (\mathbf{e}-\mathbf{b})
             =o(N^{1-1/(2m)}).
$$
But the second matrix is skew symmetric (see \eqref{G11tilde} et seq.).
By \eqref{T-16-1} the above relation
 becomes
\be\la{eq42.p}
   \mathbf{e}^T\cdot RD_{21}R\cdot \mathbf{b} +
         \mathbf{e}^T\cdot G_{11}\cdot \mathbf{b} =o(N^{1-1/(2m)}).
\ee
But from \eqref{eq31.p1}
$$
       R D_{21}R
            = - D_{12} +o(N^{1-1/(2m)})
$$
and hence \eqref{eq42.p}, \eqref{T-16-000} imply \eqref{T-15-1}.

Collecting the estimates \eqref{T-13-2}, \eqref{T-15-1l}
we see that since $\int_{-\infty}^\infty\Ai(t)\,dt=1$,
the correction term in the $11$ entry has the form
\be\la{corr11beta1}
\ba
    \frac12 \Ai(\xi) &\bigg(1-\int_\eta^\infty\Ai(t)\,dt\bigg)
        + o(1)e^{-c\xi} \\ 
     &= \frac12 \Ai(\xi) \int_{-\infty}^\eta\Ai(t)\,dt
        + o(1)e^{-c\xi} 
\ea
\ee
uniformly for $\xi,\eta\geq L_0$.
 The correction term in the $22$ entry has the same asymptotic form
 with $\xi$ and $\eta$ interchanged.
\subsubsection{The contribution of the correction term to
the $21$ entry of $K_{N,1}$}
By \eqref{M.tmp0}, \eqref{T-12p-1} the correction term in this case
has the form
$$
       \bigg(\int_x^y \Phi_1(t)^T\,dt\bigg) \cdot G_{11} \cdot\eps\Phi_1(y)
       + \bigg(\int_x^y \Phi_1(t)^T\,dt\bigg)  \cdot G_{12} \cdot\eps\Phi_2(y)
$$
which equals
$$
\ba
     \bigg(\int_x^y &\Phi_1(t)^T\,dt\bigg) \cdot G_{11}
         \cdot  \bigg(-\int_y^\infty \Phi_1(t)\,dt\bigg)\\
       &\qquad\qquad+ \bigg(\int_x^y \Phi_1(t)^T\,dt\bigg)  \cdot G_{12}
            \cdot\bigg(-\int_y^\infty \Phi_2(t)\,dt\bigg)\\
      &+\bigg(\int_x^y \Phi_1(t)^T\,dt\bigg) \cdot G_{11} \cdot\eps\Phi_1(+\infty)
       + \bigg(\int_x^y \Phi_1(t)^T\,dt\bigg)  \cdot G_{12} \cdot\eps\Phi_2(+\infty)
\ea
$$
by \eqref{T-1-00}.
Again set $x=\xin$, $y=\etan$.
A calculation very similar to the one in Subsection \ref{ssec4.3.2},
using the last estimate in \eqref{PHIsumm} in place of the first,
leads to the following asymptotic form for the $21$ correction
as $N\to\infty$
\be\la{T-25-0}
    -      \frac12 \int_\xi^\eta\Ai(s)\,ds
   +   \frac12 \bigg(\int_\xi^\eta \Ai(s)\,ds\bigg)
                     \bigg( \int_\eta^\infty \Ai(t)\,dt\bigg) +
      o(1)e^{-c\min(\xi,\eta)}
\ee
uniformly for $\xi,\eta \geq L_0$.
(Recall that there is no overall scaling factor for the $21$ entry.)
This completes the analysis of the contribution of the correction term to 
the $21$ entry of $K_{N,1}$.
\bibliographystyle{plain}

\begin{thebibliography}{0000}

\bibitem[AbSt]{Stegun} M.~Abramowitz and I.~A.~Stegun,
\textit{Handbook of Mathematical Functions, with Formulas, Graphs
and Mathematical Tables.\ }Dover, New York, 1966.

\bibitem[AFNvM]{AFNvM}
M. Adler,  P. J. Forrester, T. Nagao and P. van Moerbeke,
{\it Classical skew orthogonal polynomials and random matrices.\ }J. Statist. Phys. {\bf 99}  (2000),  141--170.

\bibitem[AvM]{AvM} M.~Adler and P.~van Moerbeke,
{\em Toda versus Pfaff lattice and related polynomials.\ }Duke Math. J.  {\bfseries 112}  (2002),  1--58.

\bibitem[Be]{Be} C.~W.~J.~Beenakker,
\textit{Universality for Br\'ezin and Zee's spectral correlator.\ }Nuclear
Phys.~B  {\bf 422}  (1994), 515--520.

\bibitem[BI]{BI} P.~Bleher and A.~Its,
{\it Semiclassical asymptotics of orthogonal polynomials,
Riemann-Hilbert problem, and universality in the matrix model.\ }Ann.~of
Math.~(2)  {\bf 150}  (1999),  185--266.

\bibitem[BoBr]{BBr} M. J. Bowick and E.~Br\'ezin,
\textit{Universal scaling of the tail
of the density of eigenvalues in random matrix models.\ }Phys. Lett. B
{\bf 268}  (1991), 21--28.

\bibitem[BrZ]{BZ} E.~Br\'ezin and A.~Zee,
\textit{Universality of the correlations between eigenvalues of large random
matrices.\ }Nuclear Phys.~B  {\bf 402}  (1993), 613--627.

\bibitem[CKu]{CKu} T. Claeys and A. B. J. Kuijlaars,
\textit{Universality of the double scaling limit in random
 matrix models,\ }preprint, 2005.
{\tt www.arxiv.org/abs/math-ph/0501074}

\bibitem[DG]{DG} P. Deift and D.~Gioev,
\textit{Universality in Random Matrix Theory for
for orthogonal and symplectic ensembles,\ }
submitted, 2004.
{\tt www.arxiv.org/abs/math-ph/0411075}

\bibitem[DGKV]{DGKV} P. Deift, D.~Gioev, T. Kriecherbauer
and M. Vanlessen,
\textit{Universality for
orthogonal and symplectic ensembles of random matrices with generalized
Laguerre type weights,\ }in preparation, 2005.

\bibitem[DKMVZ1]{DKMVZ}
P. Deift, T. Kriecherbauer, K. T.-R. McLaughlin, S. Venakides
and X. Zhou,
\textit{Uniform asymptotics for polynomials
  orthogonal with respect to varying exponential weights
  and applications to universality
 questions in random matrix theory.\ }Comm.~Pure Appl.~Math.
\textbf{52} (1999), 1335--1425.

\bibitem[DKMVZ2]{DKMVZ2}
P. Deift, T. Kriecherbauer, K. T.-R. McLaughlin, S. Venakides
and X. Zhou,
\textit{Strong asymptotics of orthogonal polynomials
with respect to exponential weights.\ }Comm. Pure
Appl. Math.~{\bf 52} (1999), 1491--1552.

\bibitem[DVZ]{DVZ} P. Deift, S. Venakides and X. Zhou,
\textit{New results in small dispersion KdV by an extension
of the steepest descent method for Riemann-Hilbert problems.\ }Int. Math. Res.
Not. \textbf{1997} (1997), 286--299.

\bibitem[DZ]{DZ} P. Deift and X. Zhou, \textit{A steepest descent
method for oscillatory
Riemann-Hilbert problems. Asymptotics for the MKdV equation.\ }Ann. of Math. (2) \textbf{137} (1993), 295--368.

\bibitem[Dy]{Dy70} F.~J.~Dyson,
{\it A note on correlations between eigenvalues of a random matrix.\ }Comm.
Math. Phys. {\bf 19} (1970), 235--250.

\bibitem[FoIKi]{FoIKi} A. S. Fokas, A. R. Its and A.V. Kitaev,
\textit{Discrete Painleve equations and their appearance in quantum gravity.\ }Comm. Math. Phys. \textbf{142} (1991), 313--344.

\bibitem[F]{F1} P.~J.~Forrester,
\textit{The spectrum edge of random matrix ensembles.\ }Nuclear Phys.~B
\textbf{402} (1993), 709--728.

\bibitem[FNH]{FNH}
P. J. Forrester, T. Nagao and G. Honner, \textit{Correlations
for the orthogonal--unitary and symplectic--unitary transitions
at the hard and soft edges.\ }Nuclear Phys. B  \textbf{553}  (1999), 601--643.

\bibitem[HWe]{HW} G. Hackenbroich and H. A. Weidenm\"uller,
\textit{Universality of Random--Matrix results for
non-Gaussian ensembles.\ }Phys.
Rev. Lett. {\bf 74} (1995), 4118--4121.

\bibitem[KaFr]{KaFr} E. Kanzieper and V. Freilikher,
\textit{Universality in invariant random-matrix models:
 Existence near the soft edge.\ }Phys. Rev. E \textbf{55}
(1997), 3712--3715.

\bibitem[KV]{KV} T. Kriecherbauer and M. Vanlessen,
{\em private communication,\ }2004.

\bibitem[MaM]{MMe91}
G. Mahoux and M. L. Mehta,
{\it A method of integration over matrix variables. IV.\ }J. Physique I France {\bf 1}
(1991), 1093--1108.

\bibitem[M1]{M} M.~L.~Mehta, \textit{Random Matrices,\ }2nd Ed.,
Academic Press, San Diego, 1991.

\bibitem[M2]{Me71} M.~L.~Mehta,
{\it A note on correlations between eigenvalues of a random matrix.\ }Comm.
Math. Phys. {\bf 20} (1971), 245--250.

\bibitem[MhSa]{MhSa}
H. N. Mhaskar and E. B. Saff, {\it Extremal problems for polynomials
with exponential weights.\ }Trans. Amer. Math. Soc. {\bf 285}  (1984), 203--234.

\bibitem[Mo]{Moore} G. Moore,
\textit{Matrix models of $2D$ quantum gravity
and isomonodromy deformations.\ }Progr. Theor. Phys. Suppl. \textbf{102}
 (1990), 255--285.

\bibitem[NW]{NW} T. Nagao and M. Wadati,
{\it Correlation functions of random matrix ensembles
related to classical orthogonal polynomials, I---III.\ }J.
Phys. Soc. Japan {\bf 60} (1991), 3298--3322,
ibid. {\bf 61} (1992), 78--88 and 1910--1918.

\bibitem[PS]{PS} L.~Pastur and M.~Shcherbina,
{\it Universality of the local eigenvalue statistics
for a class of unitary invariant random
matrix ensembles.\ }J. Statist. Phys. {\bf 86}  (1997), 109--147.

\bibitem[Ra]{Ra}
E. A. Rakhmanov, {\it Asymptotic properties of
orthogonal polynomials on the real axis.\ } Mat. Sb. (N.S.)
{\bf 119(161)} (1982),
163--203, 303. (Russian).
English transl. in: Math. Sb. (N.S.) {\bf 47} (1984), 155--193.

\bibitem[ReSi]{ReSi}
M.~Reed and B.~Simon,
\textit{Methods of modern mathematical physics, IV,\ }Academic Press,
 New York--London, 1978.

\bibitem[SaTo]{SaTo} E.~B.~Saff and V.~Totik,
\textit{Logarithmic Potentials with External Fields,\ }Grundlehren
der Mathematischen Wissenschaften, 316.
Springer--Verlag, Berlin, 1997.

\bibitem[SeVe]{SV} M. K. Sener and J. J. M. Verbaarschot,
{\it Universality in chiral random matrix theory
at $\beta=1$ and $\beta=4$.\ }Phys.
Rev. Lett. {\bf 81}  (1998), 248--251.

\bibitem[Si]{Simon} B.~Simon,
\textit{Trace Ideals and Their Applications,\ }London
Mathematical Society Lecture Notes Series, 35.
Cambridge University Press, Cambridge--New York, 1979.

\bibitem[So]{So} A. Soshnikov,
\textit{Universality at the edge of the spectrum in Wigner random
matrices.\ }Comm. Math. Phys. \textbf{207} (1999), 697--733.

\bibitem[St1]{St1} A. Stojanovic,
{\it Des probl\`emes asymptotiques dans la th\'eorie spectrale
des matrices al\'eatoires.\ }Ph.D. Thesis, Universit\'e Paris 7 Denis Diderot, 2003.

\bibitem[St2]{St2} A. Stojanovic,
{\it Universality in orthogonal and symplectic
invariant matrix models with quartic potential.\ }Math. Phys. Anal. Geom.
{\bf 3} (2000), 339--373 (2001). {\it Errata:\ }ibid., in press, 2004.

\bibitem[St3]{St3} A. Stojanovic,
{\it Universalit\'e pour des mod\`eles matriciels \`a sym\'etrie
orthogonale ou symplectique et \`a potentiel quartique.} Preprint 02-07-098,
revised version of Preprint 00-01-06.
{\tt www.physik.uni-bielefeld.de/bibos/preprints}

\bibitem[Sz]{Sz} G.~Szeg\"o,
\textit{Orthogonal Polynomials,\ }Amer. Math. Soc. Colloq. Publ., v. 23.
Amer. Math. Soc., New York, 1939.

\bibitem[TW1]{TW1} C.~A.~Tracy and H.~Widom,
{\it Fredholm determinants, differential equations and matrix models.\ }Comm.
Math. Phys. {\bf 163}  (1994),  33--72.

\bibitem[TW2]{TW2} C.~A.~Tracy and H.~Widom,
{\it Correlation functions, cluster functions,
and spacing distributions for random matrices.\ }J. Statist. Phys.
{\bf 92} (1998), 809--835.

\bibitem[TW3]{TW3} C.~A.~Tracy and H.~Widom,
{\it Matrix kernels for the Gaussian orthogonal and symplectic ensembles,\ }2004.
{\tt www.arxiv.org/abs/math-ph/0405035}

\bibitem[TW4]{TW6} C.~A.~Tracy and H.~Widom,
{\it Level-spacing distributions and the Airy kernel.\ }Comm. Math. Phys.
{\bf 159} (1994), 151--174, announcement in:
Phys. Lett. B {\bf 305} (1993), 115--118.

\bibitem[TW5]{TW4} C.~A.~Tracy and H.~Widom,
{\it On orthogonal and symplectic matrix ensembles.\ }Comm.
Math. Phys. {\bf 177} (1996), 727--754.

\bibitem[TW6]{TW5} C.~A.~Tracy and H.~Widom,
{\it Distribution functions for largest
eigenvalues and their applications.\ }Proceedings
 of the International Congress of
 Mathematicians, Vol. I (Beijing, 2002),
 587--596, Higher Ed. Press, Beijing, 2002.

\bibitem[V]{Van} M. Vanlessen,
\textit{Strong asymptotics of Laguerre-type
 orthogonal polynomials and applications
 in random matrix theory,\ }preprint, 2005.
{\tt www.arxiv.org/abs/math.CA/0504604}

\bibitem[Ve]{V} J.~Verbaarschot, \textit{Topics in
Random Matrix Theory,\ }lecture notes.
{\tt tonic.physics.\linebreak[0]sunysb.edu/\~{\space}verbaarschot/lecture/}

\bibitem[W]{W} H.~Widom,
{\it On the relation between orthogonal, symplectic
and unitary matrix ensembles.\ }J. Statist. Phys. {\bf 94} (1999),  347--363.

\end{thebibliography}

\end{document}